\documentclass[manuscript,screen,nonacm=true]{acmart}

\setcopyright{cc}
\setcctype{by}
\acmJournal{TOSEM}
\acmYear{2025} \acmVolume{1} \acmNumber{1} \acmArticle{1} \acmMonth{1} \acmDOI{10.1145/3771922}

\usepackage{subcaption} %
\usepackage{xspace}     %
\usepackage[textsize=tiny]{todonotes}
\usepackage{listings}
\usepackage{xcolor}
\definecolor{aggreen}{rgb}{0.0, 0.8, 0.6}
\usepackage[inline,shortlabels]{enumitem}

\newcommand{\head}[1]{\par\noindent\textbf{#1}~\xspace}

\newcommand{\best}[1]{\textbf{#1}}
\newcommand{\secbest}[1]{#1} %
\newcommand{\appr}{\begingroup\raisebox{1pt}{\scriptsize\ensuremath{\sim}}\endgroup}
\newcommand{\smalltt}[1]{\texttt{\small#1}}

\newenvironment{DIFnomarkup}{}{}

\usepackage{fancybox}
\usepackage[framemethod=tikz]{mdframed}
\definecolor{lightgray}{gray}{0.9}
\mdfdefinestyle{mystyle}{%
  rightline=false,%
  innerleftmargin=5,
  innerrightmargin=5,
  outerlinewidth=1pt,
  topline=false,
  leftline=true,
  bottomline=false,
  skipabove=\topsep,
  skipbelow=\topsep,
  backgroundcolor=lightgray
}

\usepackage{multirow} %
\usepackage{makecell}

\begin{document}

\title[Automated Program Repair with Round-Trip Translation]{Assessing the Latent Automated Program Repair Capabilities of Large Language Models using Round-Trip Translation}
\titlenote{Accepted for publication in ACM Trans. Softw. Eng. Methodol., \url{https://doi.org/10.1145/3771922}.}

\newcommand{\SimulaAffiliation}{\affiliation{%
  \institution{Simula Research Laboratory}%
  \city{Oslo}%
  \country{Norway}%
}}

\author{Fernando Vallecillos Ruiz}
\email{fernando@simula.no}
\orcid{0000-0001-7213-3732}
\SimulaAffiliation{}
\author{Anastasiia Grishina}
\orcid{0000-0003-3139-0200}
\email{anastasiia@simula.no}
\SimulaAffiliation{}
\author{Max Hort}
\orcid{0000-0001-8684-5909}
\email{maxh@simula.no}
\SimulaAffiliation{}
\author{Leon Moonen}
\orcid{0000-0002-1761-6771}
\email{leon.moonen@computer.org}
\SimulaAffiliation{}
\authornote{~Corresponding author.}

\begin{abstract} %

Research shows that errors in natural language can be corrected by translating texts to another language and back using language models.
We explore to what extent this \emph{latent} correction capability extends to Automated Program Repair (APR) by investigating Round-Trip Translation (RTT): translating code from one programming language into another programming or natural language and back, using Large Language Models (LLMs).
We hypothesize that RTT restores patterns most commonly seen in the LLM's training corpora through \emph{regression toward the mean}, replacing infrequent bugs with more frequent, \emph{natural}, bug-free code.
To test this hypothesis, we employ nine LLMs and four common APR benchmarks in Java, and perform a detailed quantitative and qualitative analysis of RTT-generated patches.
We find that RTT through English generates plausible patches for 100 of 164 bugs with GPT-4 on the HumanEval-Java benchmark, and 97 are found to be correct in our manual assessment. 
Moreover, RTT uniquely generates plausible patches for 
46 bugs that were missed by LLMs specifically fine-tuned for APR.
While this demonstrates the viability of RTT for APR, we also observe limitations, such as 
a lower overall bug fix rate than the state-of-the-art and diluting the original coding style.
We analyze the impact of these limitations and discuss the potential of using RTT as a complementary component in APR frameworks.

\end{abstract}



\received{22 March 2024}
\received[revised]{22 April 2025}
\received[revised]{7 October 2025}
\received[accepted]{9 October 2025}

\maketitle

\section{Introduction}

As software becomes ubiquitous and more people engage in software engineering (SE) tasks, 
the need to ensure its reliability and integrity increases. 
In the meantime, code maintenance and refactoring are tedious tasks that take a significant amount of developers' time and impede their progress in building new features~\cite{murphy-hill2015:design, bohme2017:where}.
Automated program repair (APR) aims to fix errors in source code with minimal human involvement, 
thus reducing code maintenance needs and releasing resources for creative code writing. 
With the advent of language models trained on source code, 
learning-based methods that use generative and translation models to fix bugs have started to compete with traditional heuristic and constraint-based approaches for APR~\cite{monperrus2018:automatic, legoues2019:automated}. 

Large Language Models (LLMs) trained on vast amounts of natural language and source code have pushed many fields away from traditional techniques and common heuristics, including the field of software engineering. 
Nowadays, LLMs are able to generate code, create documentation, and locate bugs~\cite{chen2021:evaluating}.
They have also been empirically proven to possess latent capabilities.
In other words, they are capable of solving tasks that they were not specifically pre-trained or fine-tuned on~\cite{bubeck2023:sparks}.

Inspired by the use of translation to find and correct errors in natural language~\cite{desilets2009:using},
we explore to what extent LLMs trained for understanding and generating code have the latent capability to find and correct errors in code. 
Specifically, we investigate an LLM's capability to debug code through a process known as \emph{round-trip translation} (RTT).
RTT was previously used to evaluate machine translation of natural languages, and more recently, 
to evaluate the correctness of code generation~\cite{allamanis2024:unsupervised}.
Unlike more traditional LLM-based APR approaches that center on one-shot bug-to-fix translations, 
RTT consists of two steps: the translation of buggy code into an intermediate language, and then the translation back to its original language. 
Thereby, RTT may offer a novel approach to repairing coding errors based on the latent error correcting capabilities of LLMs, 
an area previously overlooked by other learning-based APR techniques.

Our hypothesis is that RTT may be capable of fixing bugs as a result of a \emph{regression toward the mean} 
phenomenon exhibited by generative language models trained on vast code corpora.
Studies show that frequent code patterns in such large code corpora are bug-free~\cite{ray2016:naturalness}.
Thus, as a result of training LLMs on these corpora, 
RTT will regress toward the same mean of bug-free code, in other words, demonstrate the latent capability of producing code without errors.
To empirically investigate this hypothesis and assess the viability of RTT for APR, 
we conduct the first comprehensive study on RTT with LLMs for APR.
Our experiments use nine LLMs, including 
three GPT versions, and four APR benchmarks. 
The models vary in size, ranging from 140M 
parameters to recent OpenAI models with undisclosed sizes that have been estimated in the trillion-parameter range.
Moreover, they vary in original training objectives, spanning from code and docstring infilling to code summarization, translation, and generation. 
The benchmarks contain code with different context size and bug complexity, 
ranging from student assignments 
in QuixBugs~\cite{lin2017:quixbugs} 
and HumanEval-Java~\cite{jiang2023:impact} 
to real-world projects in Defects4J v1.2 and v2.0~\cite{just2014:defects4j}. 
Note that HumanEval-Java was not available during training of most the LLMs used in this study, 
mitigating bias from data leakage on the results obtained with this benchmark.

\begin{figure*}[b]
    \includegraphics[width=\textwidth]{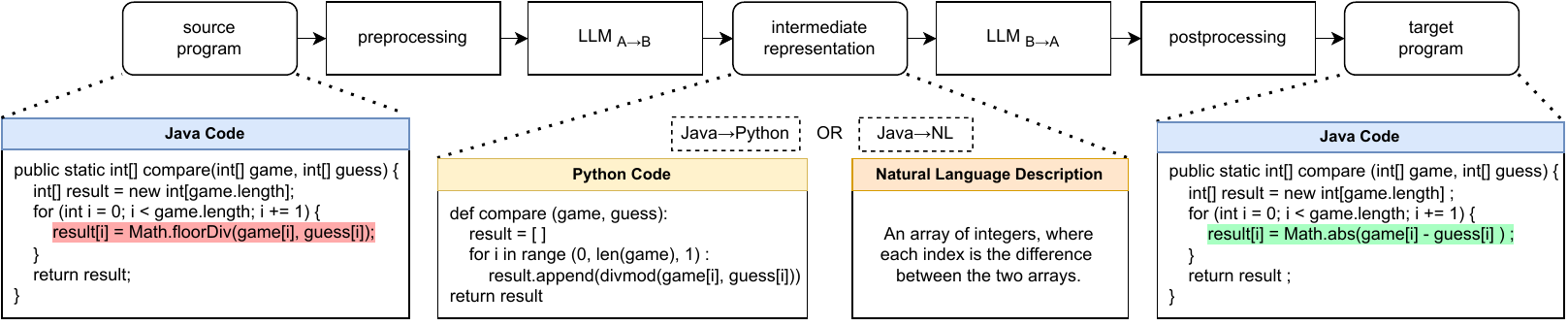}
    \caption{High-level overview of the RTT process with concrete examples taken from our empirical evaluation. The red highlight on the left indicates the buggy line, the green highlight on the right is the repaired line.}
    \label{fig:diag-rtt-process}
    \Description[RTT system overview showing translation-based bug repair using intermediate representations.]{The figure illustrates the Round-Trip Translation (RTT) process for automated program repair, transforming a buggy Java source program into a repaired target program via an intermediate representation. 
    The pipeline includes preprocessing, translation by a large language model from Java to either Python or a natural language description, followed by reverse translation back to Java and postprocessing. 
    The Java input contains a bug where integer division using Math.floorDiv is incorrect. 
    The intermediate Python code and natural language description both express the intended semantics correctly. 
    The final output shows the corrected Java code, where the division is replaced by the absolute value of the difference, using Math.abs. 
    Red and green highlights mark the buggy and fixed lines respectively, demonstrating the effectiveness of RTT in producing semantically accurate repairs.}
\end{figure*} 

In this paper, we utilize an RTT pipeline, shown in Figure~\ref{fig:diag-rtt-process}, to systematically explore the latent program repair capabilities of LLMs trained on coding tasks other than APR.
We use LLMs initially trained or fine-tuned for code translation, summarization, and generation, 
to translate buggy code from one language to another and back, investigating whether this process implicitly fixes the bugs. 
We assess how effectively this approach eliminates bugs under various conditions.
The pipeline uses either a programming language (PL) or a natural language (NL) as intermediate representation. 
Moreover, our RTT pipeline uses the LLMs in a zero-shot fashion: 
unlike other neural APR methods, 
with RTT we do not need to fine-tune models on the bug repair task, 
but apply them \emph{off-the-shelf}, as provided by the model authors. 
Our investigation aims to understand the characteristics of RTT-based repair, 
including its potential strengths and weaknesses compared to direct bug-fixing approaches. 
A key aspect of our analysis involves determining if RTT can address bugs that are missed by other methods, 
which could inform its potential role within the broader APR landscape.
Our findings suggest the potential of RTT as a complementary tool in a collaborative LLM-based APR framework,
where different models and their pipelines generate and update code.

\head{Contributions:} The main contributions of this work are as follows: 
\begin{enumerate}[$\star$,nosep]
\item We present the first systematic empirical evaluation of RTT with LLMs for APR, 
an area previously overlooked by other APR approaches;
\item We asess RTTs effectiveness across nine language models (six open-source models and three API-based), 
four common APR benchmarks with various context sizes and bug types, and 10 different seeds; %
\item We compare the effectiveness of using another programming language versus using natural language as the intermediate representation in the RTT pipeline; 
\item We analyze the trade-offs between LLM size, temperature, and repair performance; 
\item Our findings include that RTT generates plausible patches for 100 of 164 bugs in the HumanEval-Java benchmark, and 97 were found to be correct in our manual assessment. 
Moreover, over all four benchmarks, RTT generates plausible patches for 46 bugs that were not fixed by other methods, even those fine-tuned on the APR task. 
Note that the evaluation of patches for QuixBugs and Defects4J only considered patch plausibility, which has inherent limitations, such as ignoring potential overfitting to test code; 
\item We provide an in-depth qualitative analysis of RTT-generated patches for QuixBugs and HumanEval-Java, identifying common repair patterns, characteristics, and underlying challenges associated with the approach;
\item We conclude by discussing the benefits, limitations, and potential role of RTT in the APR landscape, based on our empirical findings;
\item We release the code for RTT and results obtained to ensure replication and verification of our work, 
including the manual assessment of patch correctness on over 5,000 RTT-generated patches for HumanEval-Java.\footnote{~\label{replication}%
The replication package is available for download from Zenodo: \url{https://doi.org/10.5281/zenodo.10500593}.
}
\end{enumerate}

\section{Background and Related Work}
\subsection{Neural Machine Translation}
Traditional methods for translating text from one language into another are increasingly replaced by Neural Machine Translation (NMT), where neural networks are used to predict a sequence of translated words~\cite{stahlberg2020:neural,yang2020:survey} and sequence-to-sequence methods, in general~\cite{sutskever2014:sequence}. 
Sequence-to-sequence models 
enable NMT to 
automatically learn complex mappings between different languages, efficiently capturing context and offering more accurate translations in comparison to their predecessors.

To generate translated sentences, autoregressive NMT methods use the whole source sentence and the initial part, or \emph{prefix}, of the target sentence.
Assuming $x =\left\{x_1, ..., x_n\right\}$ is the source sentence
split into $n$ tokens, $y=\left\{y_1, ..., y_m \right\}$ is the target sentence and 
$y_{<i} = \left\{y_1,...,y_{i-1} |\ i \leq m \right\} $ 
is the beginning of the target sequence generated up to token $i-1$,
we can formulate the generation as a conditional probability $P(y|x)$:
\begin{equation}
\label{eq:nmt_translation_probability}
    P(y|x) = \prod_{i=1}^{m}P(y_i |x, y_{<i}).
\end{equation}

Notably, NMT methods are applied to translate between both natural and programming languages. 
The predominant model architecture is encoder-decoder or decoder-only transformer~\cite{vaswani2017:attention}.
A widely used approach to create an NMT model for PL-PL translation is to fine-tune a model pre-trained on code and, possibly, natural language on a dataset with pairs of code snippets in the source and target PLs~\cite{ahmad2021:unified, wang2021:codet5,fried2023:incoder}.
Similarly, GPT models by OpenAI are sequence-to-sequence models trained to create an output text or code sequence (response) from an input sequence (prompt), the latest models benefiting from Reinforcement Learning with Human Feedback (RLHF) and other training advancements as well as large amounts of pre-training data~\cite{radford2019:language, brown2020:language, openai2023:gpt4}.
Overall, the incorporation of deep learning-based NLP techniques left a distinct mark on the field of data-driven software engineering methods by allowing it to leverage statistical properties such as code naturalness.

\subsection{Software Naturalness}
\emph{Software naturalness} is the property of source code to exhibit patterns and follow conventions that are statistically similar to other forms of human expression, such as natural language~\cite{hindle2012:naturalness, allamanis2018:survey}. 
Due to the repetitive nature of coding patterns and the natural incline of software developers to create simple blocks of code for readability and maintainability, programming languages have learnable statistical properties~\cite{hindle2012:naturalness}.
This means that NLP techniques that rely on learning patterns in input sequences, such as NMT, can be applied to source code. 
Neural Program Translation (NPT) applies NMT to understand the underlying logic and semantics of the source code and generates functionally equivalent programs in the target language~\cite{roziere2020:unsupervised}.
Ray et al.~\cite{ray2016:naturalness} observed bugs to be deviations that manifest as unnatural \emph{noise} which increases entropy in otherwise predictable and repetitive natural code. 
This observation has been used to address various tasks, such as APR~\cite{tufano2019:empirical}, vulnerability identification~\cite{buratti2020:exploring}, and patch ranking~\cite{kang2022:language, kolak2022:patch}.

\subsection{Language Models for Automated Program Repair}

Interpreting APR as a translation from buggy to fixed sequences further stimulated the use of language models~\cite{chen2021:sequencer, jiang2021:cure}.
Early models used RNN and LSTM architectures that cannot handle long-range dependencies and scale poorly~\cite{zhang2023:survey}.
Transformers~\cite{vaswani2017:attention} addressed these challenges and are now the prevalent choice. 
Transformers for APR and other SE tasks have evolved by
incorporating new representations~\cite{guo2021:graphcodebert},
new loss functions~\cite{jain2021:contrastive}, 
and increasing their size~\cite{xia2023:automated}.  
This enables them to understand more complex syntactical structures~\cite{niu2022:deep, zeng2022:extensive}, 
and make human-competitive repairs~\cite{fan2023:automated}. 

Recent LLM-based approaches to code repair include Neural Machine Translation (NMT)~\cite{chen2021:sequencer}, cloze-style repair~\cite{xia2022:less}, instruction and chat-based methods, and frameworks with iterative LLM calls~\cite{shinn2023:reflexion, chen2023:teaching}.
NMT methods are trained to repair code by translating from buggy lines to fixed lines, i.e., by learning repairing edits. 
In the cloze-style repair approach, buggy lines are masked to allow a model to infer correct code directly given surrounding code tokens as context. 
Recent work has leveraged the use of zero-shot LLMs to successfully perform cloze-style APR on numerous benchmarks~\cite{xia2022:less, xia2023:automated,jiang2023:impact}.
Instruction-based LLMs and chat-based LLMs are adept at following the queries specified by the user.
In the context of APR, developers may describe the bug or the intended behaviour of the code snippet.
The LLM would suggest precise fixes or generate corrected code snippets directly~\cite{xia2023:keep, xia2023:conversational}.
Chat-based LLMs add a layer of accessibility to APR which 
developers are able to engage with.
The iterative process of refining queries based on feedback allows a deeper understanding of the problem and more customized solutions~\cite{chen2023:teaching, sobania2023:analysis}.

Frameworks with iterative debugging use self-reflection abilities of language models to provide feedback to past partial solutions and debug the code further. 
Meanwhile, combining different LLMs and approaches to one framework via autonomous agents has been a recent tendency~\cite{yahav2023:aidriven}. 
Therefore, the code repair task, similar to other automated decision-making tasks, is prone to be solved with a collection of tools rather than one model. 
In this line of thought, round-trip translation is a novel approach to repairing code by translating it to an intermediate language and then back. 
While in cloze-style and NMT settings, a patch is generated in one single attempt, 
and in iterative debugging~\cite{grishina2025:fully}, a patch is evolved by a number of calls to the LLM, 
RTT attempts to repair the code in exactly two steps.
Moreover, NMT and cloze-style techniques differ from the RTT approach proposed in this paper in that they are fine-tuned on, or prompted to perform, the APR task, whereas the proposed RTT approach uses LLMs without any fine-tuning or prompting for APR.

\subsection{Round-Trip Translation}
RTT involves translating a text from its original language to an intermediate language and then translating it back to the original language. 
Our use of RTT was inspired by the practical observation that, for our secondary languages, 
we would check or correct errors using RTT through publicly available NMT tools. 
The value of this practice was confirmed in a study by Hermet and D{\'e}silets~\cite{desilets2009:using}.
Other uses of RTT in NMT include improving translation results~\cite{mehta2020:simplifythentranslate}, 
and testing the accuracy of a translation model~\cite{zhuo2023:rethinking}.
In the context of APR research, RTT has previously been used for data augmentation~\cite{ahmad2023:summarize,silva2023:mufin} and evaluating code correctness~\cite{allamanis2024:unsupervised}.

\section{Repair through Round-Trip Translation}

\noindent
Our study evaluates the effectiveness of APR based on round-trip translation (RTT) using nine recent LLMs.
A high-level overview of the RTT approach was presented earlier in Figure~\ref{fig:diag-rtt-process}, 
illustrated with concrete examples that are taken from our experiments.
Specifically, the process uses LLMs for two subsequent translations: 
first, to translate buggy code from its source language to an intermediate language 
(which can be either another programming language or a natural language), 
and second, to translate the intermediate representation back to the original source language.

\subsection{Motivation to Use Round-Trip Translation for Program Repair}
We hypothesize that RTT is capable of repairing bugs as a result of the \emph{regression toward the mean} or homogenization phenomenon exhibited by generative language models.
The reasoning is as follows: The LLMs employed in RTT are trained on vast real-world code corpora. 
Such models can treat code as natural language due to the naturalness hypothesis~\cite{hindle2012:naturalness}. 
They generate or summarize code by iteratively selecting the sequence of the most probable tokens, or the most probable sub-sequences of tokens, for example, using beam search~\cite{sutskever2014:sequence}.
The probability is estimated by the language model based on its weights, and adapted during the model training to return the most frequently occurring tokens in similar contexts. 
Ray et al.~\cite{ray2016:naturalness} have shown that frequent code patterns in large real-word code corpora are bug-free.
Thus, as a result of training LLMs on these corpora, they 
have a tendency to generate code that is also bug-free.
This process in which LLMs return the most probable tokens during generation can be viewed as regression toward the mean, where noisy samples are replaced by samples closer to the mean.
Therefore, each translation step in RTT homogenizes the source fragment toward a less noisy variant that is closer to the expected most probable code, a patch candidate. 
In the context of code, bugs have been shown to act as a form of noise that is less natural than the mean~\cite{ray2016:naturalness}, so they should be reduced or eliminated over the course of round-trip translation, thereby making APR a latent capability of LLMs trained on code.

\subsection{Formulation of Round-Trip Translation}

Formally, our approach can be described as follows. 
Let $x = \left\{x_1, ..., x_n\right\}$ be a buggy code snippet split into $n$ tokens and 
$\tilde{x} = \left\{\tilde{x}_1, ..., \tilde{x}_m\right\}$ 
its round-trip translated version with $m$ tokens, i.e., a candidate patch. 
We use LLMs as neural machine translation models: $LLM_{A \rightarrow B}(\cdot)$ from language $A$ to $B$, where $A \neq B$, and $LLM_{B \rightarrow A}(\cdot)$.
The round-trip translation of a code snippet $x$ is a two-legged translation.
The first leg, \emph{forward translation}, produces a sequence in language $B$, and the second one, \emph{backward translation}, generates code in language $A$ from the sequence in language $B$. 
The whole process can be expressed as:
\begin{equation}
\label{equ:rtt-a-b-a} 
    \tilde{x} = LLM_{B \rightarrow A }(LLM_{A \rightarrow B}(x)).
\end{equation}

The total probability of the candidate patch generated by RTT, can be expressed with an intermediate representation $r$ of $x$ as follows:
\begin{equation}
    \label{equ:total-probability}
    P \left( \tilde{x} \right) = 
    P\left( \tilde{x} | r \right) \cdot P(r).
\end{equation}
Probabilities $P(r)$ and $P\left( \tilde{x} | r \right)$ can be approximated with available LLMs according to Eq.~\ref{eq:nmt_translation_probability}. 
Therefore, we use two legs of translation to approximate the candidate patch 
$\tilde{x}$ 
in Eq.~\ref{equ:total-probability}:
\begin{eqnarray*}
    P \left( \tilde{x} \right) 
    = 
    P(r) \cdot P\left( \tilde{x} | r \right) 
    \approx 
    \prod_{i=1}^{k}P_{LLM_{A \rightarrow B}}(r_i |x, r_{j<i}) 
    \cdot
    \prod_{i=1}^{m}P_{LLM_{B \rightarrow A}}(\tilde{x}_i |r, \tilde{x}_{j<i}).
\end{eqnarray*}
In this work, we use different intermediate languages $B$ with the goal of encouraging 
a diverse range of representations, 
namely natural language (English) and programming languages.

We also formalize the notion used to investigate if RTT can indeed repair bugs.
We denote a benchmark with $N$ buggy code snippets as $ \{ x^i \}_{i=1}^N$, 
and let $Plausible(x) \rightarrow \left\{0; 1\right\}$ be a function that returns $1$ if code snippet $x$ passes all test cases~\cite{zhang2023:survey}. 
Then, to evaluate if RTT can indeed repair bugs, we 
check if a collection of snippets after round-trip translation has a higher overall plausibility than the original collection, expressed by the following equation:
\begin{equation}\label{eq:plausible}
    \sum_{i=1}^{N}Plausible(\tilde{x}^i) > \sum_{i=1}^{N}Plausible(x^i).   
\end{equation}
The practical implementation and evaluation of RTT is discussed in more detail in Section~\ref{sec:implementation}.

\section{Experiment Design}
\label{sec:experiment}

The following five research questions guide our evaluation of RTT's performance on the APR task:
\begin{enumerate}[label=\textbf{RQ\arabic*:},nosep]
\item How well does RTT perform repairs with a programming language as intermediate representation? 
\item How well does RTT perform repairs with natural language (specifically English) as intermediate representation?
\item How sensitive is RTT repair performance to variation of the LLM's temperature hyperparameter?
\item What quantitative trends can be observed in the patches generated by RTT?
\item What qualitative trends can be observed in the patches generated by RTT?
\end{enumerate}
To address these research questions, we use nine LLMs and four APR benchmarks discussed in detail below.
The selection of these models and benchmarks was guided by ensuring a diverse and thorough evaluation of RTT for APR.

\subsection{Models}
\label{sec:models}
We use 
nine
distinct transformer-based language models for our evaluation. 
Their sizes, architectures, and characteristics of training datasets are shown in Table~\ref{tab:model-sizes}. 
We select the models based on two main requirements: 
\emph{(i)} they are trained on large code corpora and perform well on code-related tasks; 
\emph{(ii)} they can perform both legs of a round-trip translation, through an NL or another PL. 
None of the model variants we use were originally trained or fine-tuned for code repair, and we use them \emph{as-provided}, 
without additional fine-tuning or training.
Although the models' original goal was not code repair, we consider the \emph{outputs} of the backward (second) translation leg as \emph{candidate patches} in our experiments.
Observe that one can choose to use different models in each leg of the translation, removing the need for our second requirement. However, in the context of this paper, we use the same model in each leg.
This decision is discussed further in subsection \ref{subsec:further-details}.

\begin{DIFnomarkup} %
\begin{table}[b]
\centering
\newcommand{\datasize}[1]{}%
\setlength{\tabcolsep}{7pt}
\caption{Overview of language models used for RTT.}
\begin{tabular}{lcccc}
\toprule
Model & \makecell{Size} & Base Model & Architecture & \makecell{Data Source} \\ 
\midrule
{PLBART} &
  \makecell{base (140M)} &
  BART & encoder-decoder & 
  \makecell{StackOverflow \\ BigQuery \datasize{655 GB}} \\[3ex]
{CodeT5} &
  \makecell{base (220M)} &
  T5 & encoder-decoder & 
  \makecell{CodeSearchNet\\ BigQuery\\ \datasize{8.35M functions}} \\[3ex]
{TransCoder} & \appr440M & 
  T5 & encoder-decoder &  
  \makecell{Google BigQuery \datasize{744 GB}}  \\ 
\midrule
{SantaCoder} & 
  1.1B & 
  GPT-2 & decoder & 
  \makecell{The Stack (v1.1) \datasize{268 GB}} \\[3ex]
{InCoder} &
  \makecell{1.3B\\ 6.7B} &
  Mixture of Experts & decoder & 
  \makecell{StackOverflow\\ GitHub/GitLab \datasize{216 GB}} \\[3ex]
{StarCoderBase} & 
  15.5B & 
  GPT-2 & decoder & 
  \makecell{The Stack (v1.2) \\ \datasize{815 GB}} \\ 
\midrule
{GPT-3.5-Turbo\quad} & 
  Not Disclosed & 
  GPT-3 & decoder & 
  Public Data \\[3ex]
{GPT-4} & 
  Not Disclosed & 
  GPT-4 & decoder & 
  Public Data \\[3ex]
{GPT-4o-mini} & 
  Not Disclosed & 
  GPT-4o & decoder & 
  Public Data \\ 
\bottomrule
\end{tabular}%
\label{tab:model-sizes}
\end{table}
\end{DIFnomarkup} %

\head{PLBART}~\cite{ahmad2021:unified} (Programming Language Bidirectional and Auto-Regressive Transformer) is a large language model designed for code-related tasks. 
Based on the BART~\cite{lewis2020:bart} model, at the time of model release it excelled in understanding and generating code across different programming languages. 
PLBART follows the same architecture as its predecessor, a sequence-to-sequence transformer architecture with 6 encoder-decoder layers. Only one size was released initially: PLBART-base (140M).
However, a bigger version, PLBART-large (400M), was distributed  later in the same year.  

PLBART is trained on a dataset in Java, Python, and natural language (English). 
As a result, the model has learned the patterns, semantics, and syntax of programming languages and grasped the foundations of code understanding and generation. 
This results in a base model that is versatile and able to achieve decent results throughout a large range of tasks.
At the time of its publication, PLBART achieved state-of-the-art results in text-to-code, code-to-text, and code-translation tasks.
In addition, the authors have fine-tuned the model on a series of downstream tasks and released the checkpoints. 
Furthermore, they also studied and released models fine-tuned on multi-tasking corpora and multi-language which can increase the versatility and robustness of the model even further~\cite{ahmed2022:multilingual}. 
At the time of the writing of this work, the authors have released a total of 53 fine-tuned models.
Of those, we use the \emph{base} models fine-tuned on 
translation between Java and C\# and the \emph{base} models fine-tuned on code summarization (Java $\rightarrow$ NL) and code generation (NL $\rightarrow$ Java).

\head{CodeT5}~\cite{wang2021:codet5} is a large language model designed for code-related tasks. It is based on the T5 model (Text-to-Text Transfer Transformer)~\cite{raffel2020:exploring} and is further tuned for code understanding. Like its predecessor, CodeT5 has a transformer-based architecture.
Introduced by Google researchers, 
T5 was designed as a series of multiple encoder-decoder layers that capture contextual information and relationships in the code. 
To this end, CodeT5 was trained with a new identifier-aware denoising objective that aims to improve the model's code comprehension.
Initially, two sizes were released: CodeT5-small (60M) and CodeT5-base (220M). 
A year after this release, and additional was delivered: CodeT5-large (770M)~\cite{le2022:coderl}. 
At the time of the writing of this work, a new family of models has been released CodeT5+~\cite{wang2023:codet5} that includes five different sizes (220M, 770M, 2B, 6B, 16B).

CodeT5 follows a two-step training process. 
The first step results in a base model which accumulates most of the code understanding but does not excel in any task specifically. 
Fine-tuning the model toward specific code-related tasks such as summarization, translation, and generation (among others) has resulted in the model achieving its best results.
The authors have further trained and released a series of fine-tuned checkpoints that cover most of the tasks in the CodexGLUE benchmark~\cite{lu2021:codexglue}. 
In some of these tasks, the authors have also fine-tuned the model to cover multiple programming languages, resulting in a more general knowledge model while maintaining most of its fine-tuned advantage~\cite{ahmed2022:multilingual}. 
We use the \emph{base} size models fine-tuned on the same types of tasks as PLBART. 

\head{TransCoder}~\cite{roziere2020:unsupervised} is a large language model specifically created with the sole purpose of source code translation between different programming languages, namely C++, Java, and Python.
Researchers at Meta AI focused on creating a transcompiler or source-to-source translator. 
Rule-based transcompilers often require larger resources and perform worse than the neural approaches. 
However, the lack of data poses a major challenge in the code-translation domain.
TransCoder uses an unsupervised approach to overcome this obstacle and relies simply on monolingual source code.

Researchers trained the model with the three principles of unsupervised machine translation (initialization, language modeling, back-translation) as indicated by Lample et al.~\cite{lample2018:phrasebased}. 
They transformed these principles into three forms of training: cross-lingual masked language model training, denoising auto-encoding, and back-translation. 
These techniques can take place with monolingual data and, in the case of back-translation, in parallel input corpora, i.e., the corpora of aligned examples in two languages. 
This resulted in a model that outperformed every other baseline such as \textit{j2py}\footnote{https://github.com/natural/java2python} at the time by a significant margin, as well as an easily generalizable model for other programming languages.
This makes it a promising candidate for an RTT pipeline, where we aim to reduce noise or bugs in two steps, from the original buggy example to the translated intermediate representation and back to obtain the final candidate patch. 

\head{SantaCoder}~\cite{allal2023:santacoder} is an open-source large language model designed for tasks related to code. 
The model was developed by the BigCode project, a scientific collaboration whose goal is the creation of ethical and responsible large language models for code.
Currently, it supports three programming languages: Java, JavaScript, and Python.
Its structure is a decoder-only transformer and holds approximately 1.1B parameters. 

SantaCoder features multi-query attention and was trained on the Stack dataset (v1.1)~\cite{kocetkov2023:stack} through the fill-in-the-middle objective. 
Despite its relatively small size, the authors claim that the model achieves comparable to improved performance when juxtaposed with previous multilingual models such as InCoder-6.7B. 
The authors showed that SantaCoder outperformed these models in infilling and left-to-right generations for some benchmarks. 
Since SantaCoder is released under an OpenRAIL license, the support of this model also reinforces the development of ethical and responsible models for code.
We use SantaCoder 
for RTT with NL since the model learned to operate with docstrings during pre-training.

\head{InCoder}~\cite{fried2023:incoder} is a large language model able to perform program synthesis as well as code editing in 28 programming languages. 
The researchers' goal 
was to design a
model that is able to infill arbitrary regions of code in different settings such as comment generation, type inference, and variable-renaming. 
InCoder differs from its competitors by offering left-to-right generation, masking, and infilling in a single model. 
Therefore, it offers a comprehensive solution for code manipulation and generation.

The model was trained with a causal masking objective, which allows the extraction of context from both sides of the infilled region~\cite{aghajanyan2022:cm3}. 
The training corpus consisted of code files and repositories from GitHub and GitLab, as well as a corpus extracted from StackOverflow. 
The structure of the model follows the one described by \citet{artetxe2022:efficient}. 
InCoder can also perform zero-shot tasks while still achieving comparable performance to similar left-to-right code synthesis models. 
This result indicates that the additional editing and infilling capabilities of the model do not hinder its program synthesis capabilities. 
The authors released two different sizes of the model, 1.3B and 6.7B parameters, showing that although trained with the same data, the increased size also increased its performance. 
These models offered a powerful tool with the ability of code infilling using bidirectional context that expanded on previous left-to-right generational models.
We use InCoder in a similar fashion to SantaCoder. 

\head{StarCoder}~\cite{li2023:starcoder} follows the same structure as SantaCoder. The main difference is that StarCoder is a 15.5B-parameter model.
StarCoder is an open-source large language model trained on source code and natural language. 
Being also developed by the BigCode project, its creation contributes to the development of ethical and responsible language models for code. 
Furthermore, this model was trained on over 80 programming languages, Jupyter notebooks, and a large amount of Git communications such as commits or issues. 

Similar to its predecessor, StarCoder also features multi-query attention and was trained on a bigger version of the Stack dataset (v1.2)~\cite{kocetkov2023:stack} through the fill-in-the-middle objective. 
The authors claim that StarCoder outperforms every other multi-lingual open large language model for code and matches the OpenAI \textit{code-cushman-001} model. 
They released two versions of the model, StarCoderBase and StarCoder, which is a further fine-tuned version in Python. 
The model is also able to handle a context of up to 8 thousand tokens.
Longer contexts allow for new and improved applications of the model such as code autocompletion, modifications through instructions, or summarization of code. 
StarCoder is also released under an OpenRAIL license, supporting the goals of the BigCode project. 
Two versions are released: StarCoderBase and StarCoder (fine-tuned on Python), of which we use StarCoderBase in the infilling mode for experiments with NL as intermediate.

\head{GPT-3.5}~\cite{brown2020:language}, \textbf{GPT-4}~\cite{openai2023:gpt4}, and \textbf{GPT-4o-mini}~\cite{openai2024:gpt4o} 
are 
models from OpenAI's series of decoder-only transformer-based large language models. 
Unlike the previously mentioned models, they are not designed solely for code-related tasks. 
The extensive training data, which includes natural language and code, provides the necessary knowledge to perform a wide range of tasks including code-related ones. 
Despite the similarity of GPT-4 to its predecessor, GPT-3, the significant increase in parameters leads to the model outperforming in code-related and NLP tasks. 

These models perform well in tasks such as code translation, summarization, and generation, even though they were not trained or fine-tuned for code-related tasks.
The ability to handle large contexts and their general purpose natural and programming language understanding allow the models to grasp the syntax and semantics of the code in an indirect way. 
These aspects also make the models more versatile for any tasks that may require natural language and source code, such as code summarization or generation. 
The models are also the only completely closed models in this section. Therefore, any interaction with them must be done through the OpenAI API provided.\footnote{~See \url{https://platform.openai.com/docs/guides/gpt}. 
Specifically, we use the 
\textit{gpt-3.5-turbo-1106}, \textit{gpt-4-0613}, and \textit{gpt-4o-mini-2024-07-18} 
models.
}
However, the availability of API endpoints has also facilitated real-world application of the models in many fields such as support, sales, content creation, and programming among others.
We have chosen to use the GPT models only with NL as intermediate to limit the costs. 

\subsection{Benchmarks}

We have chosen four diverse APR benchmarks, following Jiang et al.~\cite{jiang2023:impact}.
In this choice, only single-hunk bugs are included.
We follow their example because it enables
direct comparison between the RTT approach and results of previous work on the NMT-style APR, in which buggy code is directly translated to patches. 
Furthermore, this decision only modifies the original size of two benchmarks: Defects4J v1.2 and Defects4J v2.0.
The other two benchmarks already contain only single-hunk bugs.
Single-hunk bugs are frequently chosen in software engineering benchmarks since their scope allows for both, quantitative metric evaluation and feasible manual analysis of patch quality.
All the benchmarks are in Java and contain buggy and fixed code, as well as tests to check the test pass rate of the candidate patches generated by RTT.
Note that we use the concepts \emph{problem}, \emph{bug}, and \emph{code example} interchangeably, because we use buggy code examples with single-hunk bugs only. 

\head{QuixBugs}~\cite{lin2017:quixbugs} is a program repair benchmark with 40 buggy common algorithmic programs such as \textit{bitcount} and \textit{bucketsort} and their fixed versions. These problems were collected in the Quixey Challenge, in which programmers were provided with a buggy implementation of a classic algorithm that would need to be fixed in under a minute. Given that the problems were designed as a programming challenge for humans, it is a 
valuable
testbed for program repair techniques given their diversity and realism.

Manual translation from Python to Java was done carefully after the challenge. Given the differences between the programming languages, many revisions took place. Special attention was put into maintaining the one-line confinement of all bugs despite Java's verbosity. Although the initial version of QuixBugs was not entirely suitable for automatic program repair in Java, recent work has thoroughly reviewed and improved the dataset by providing correct versions and integrating their findings in a newer version that is used to evaluate automatic repair techniques~\cite{ye2021:comprehensive}. This update greatly enhances the applicability of the benchmark and allows for a more comprehensive evaluation of APR techniques. Currently, the benchmark consists of 40 programs in Java and Python with one single-hunk bug in each. 

The original classification of the programs was performed based on defect types, providing a basic foundation and guide such as \textit{incorrect variable} or \textit{incorrect comparison operator}. However, posterior classifications have been done ad-hoc to further refine the understanding of the bugs~\cite{ye2021:comprehensive}. This makes QuixBugs a valuable resource to evaluate and compare automatic repair techniques. Although it consists of a relatively small number of programs, the diverse set of bugs enables comparative studies against realistic problems. Furthermore, the coverage of multiple languages allows the testing of a wider range of tools and more diverse and extensive analysis.

\head{HumanEval-Java}~\cite{jiang2023:impact} is introduced to
address the potential issue that some of the APR benchmarks may have been used as source for training data for some models.
HumanEval-Java is derived from HumanEval~\cite{chen2021:evaluating} and, unlike its predecessor, this benchmark is tailored for automatic program repair. Although the original dataset was written in Python, the authors manually translated every problem to Java along with their test suites. Furthermore, bugs were injected into the Java programs, creating the final dataset composed of 164 single-hunk Java bugs. Single-hunk bug datasets are very desirable in the field of automatic program repair since they allow users to try different techniques and models even with small context sizes while facilitating benchmarking and maintaining real-world relevance. The authors aimed to introduce a range of different defects, from simple incorrect operator usages to more complex logical bugs that may require the removal or editing of multiple lines to fix. This manual translation and bug injection ensures that no model has been trained on this benchmark before, making it an unbiased and fair dataset in Java to compare multiple models.

\head{Defects4J v1.2} is one of the first releases of the Defects4J  benchmark~\cite{just2014:defects4j}.
In its turn, Defects4J is one of the most recognizable benchmarks in the field of APR. It has built a bridge between the research and real-world problems people faced in their work. The benchmark consists of a collection of reproducible bugs from open-source Java projects. Furthermore, test suites are provided which significantly facilitates the use of iterative tools for APR. These bugs have a wide range of complexity and are related to multiple domains, offering users a comprehensive benchmark for new tools and techniques. Over time, Defects4J has been updated resulting in different versions with deprecated bugs and additional projects.

Defects4J v1.2 provides 395 bugs (4 are deprecated) from six open-source Java projects (Chart, Closure, Lang, Math, Mockito, and Time) along with different properties of the defects, the buggy and fixed version, dependencies, and triggering tests. Each defect has a different test suite provided to automate the new test.
Researchers in academia and industry have leveraged this benchmark to study different iterative techniques such as genetic algorithms and machine learning approaches. The availability of unit tests along with the detailed metadata allow thorough comparative studies among techniques, bug types, project sizes, etc. Moreover, it gave generalizability, effectiveness, and efficiency measures for new APR tools. The deeper analysis of this release~\cite{sobreira2018:dissection} along with its novelty resulted in this benchmark becoming one of the essentials in the automatic program repair field.

\head{Defects4J v2.0} is the latest stable version of the benchmark. It includes additional 438 bugs from nine open-source Java projects. Every bug comes with relevant metadata, a fixed version, and a test suite. This new version provides an even more thorough and diverse collection of defects to test APR techniques.
Since this benchmark follows the same structure as the previous versions, researchers can easily expand their experiment and confirm or disprove their former results. The inclusion of new defects and projects only enhances the reliability of the benchmark for evaluating the mentioned metrics. The inclusion of more projects is especially essential to measure the transferability of the techniques across different bug types and environments.

\subsection{Implementation of Round-Trip Translation}
\label{sec:implementation}

The RTT pipeline comprises four main steps: preprocessing of input buggy code, generation of translations, postprocessing of RTT-generated outputs, and their validation. 
We refer to \emph{input} source code as \emph{buggy code}, \emph{buggy examples} or simply \emph{bugs}, while \emph{outputs} of the RTT pipeline correspond to \emph{candidate patches} in APR terminology.
The first three steps are shown in Figure~\ref{fig:diag-rtt-process}.
We add the fourth evaluation step in the current section. 
The result is a versatile and parallelizable pipeline able to generate and validate RTT patches with diverse models and  test against different benchmarks. 
Our pipeline extends the framework of Jiang et al.~\cite{jiang2023:impact}. 

\head{Step 1: Preprocessing and Prompting.~} 
We follow the common practice and extract solely the buggy function as is also done by Jiang et al.~\cite{jiang2023:impact}. 
To conform with language models requirements, we add prefixes, suffixes, masks and/or general style changes, such as removing newline characters, before tokenization. 
We insert a Javadoc header that serves as a prompt for the infilling models (SantaCoder, StarCoderBase, InCoder). 
The extraction of code and preprocessing is automated and requires no human intervention.
Section~\ref{sec:prompt-choice} contains the exact prompts used for the models.

\head{Step 2: Round-trip Translation.~}
In the round-trip translation step, we generate two translations for each buggy example using the same type of LLM for both RTT legs: from preprocessed buggy code to an intermediate language and from the intermediate language back to the original language. 
We generate five different translations per leg in the round-trip translation using LLMs with non-zero temperature to ensure the diversity of the intermediate representations and final candidate patches.
Therefore, for each buggy example in a benchmark, we obtain five translations in the intermediate language and 25 final candidate patches, i.e., five from each intermediate translation. 
Temperature and other model-specific hyperparameters are described in Section~\ref{section:hyperparemeters}.

\head{Step 3: Postprocessing.~}
We also perform minor postprocessing of the RTT-generated candidate patches to ensure that function signatures are as expected by the test suites. 
Therefore, we extract code if both code and text are generated and remove extra tokens, which also increases the readability of patches.
This process is also automated and does not require human intervention.
Section~\ref{sec:postprocess-ensure-testability} contains more details on postprocessing with an NL and PLs as intermediate translations. 

\head{Step 4: Evaluation of RTT Results.}
The final step evaluates the postprocessed RTT results against the test suites provided by the benchmarks. 
We calculate additional metrics for each candidate patch to evaluate the performance of the models and measure the effectiveness of RTT for APR. 

\subsection{Hyperparameters for Language Models}
\label{section:hyperparemeters}
We use the recommendation of the authors of the models when choosing the hyperparameters, unless specified otherwise in the current section. 
All the final hyperparameter values are reported in Table~\ref{tab:hyper}. 
In detail, we set the number of beams to 10 and the temperature to 1 for 
PLBART,\footnote{~\url{https://github.com/wasiahmad/PLBART}}
CodeT5,\footnote{~\url{https://github.com/salesforce/CodeT5}}
and TransCoder.\footnote{~\url{https://github.com/facebookresearch/TransCoder}} 
For SantaCoder,\footnote{~\url{https://huggingface.co/bigcode/santacoder}} 
StarCoder,\footnote{~\url{https://huggingface.co/bigcode/starcoderbase}}
and InCoder,\footnote{~\url{https://huggingface.co/facebook/incoder-1B}}%
$^{,}$\footnote{~\url{https://huggingface.co/facebook/incoder-6B}}
we follow the hyperparameter setup reported in their public demos with a number of beams of 1 and Top-P nucleus sampling of 0.95. 
Although this choice is reflected in the respective publications \cite{fried2023:incoder, allal2023:santacoder, li2023:starcoder}, they chose a temperature of 0.2 when generating a single output and a temperature of 0.8 when generating up to 100 outputs.
We found that a temperature of 0.2 led to insufficient variation.
To account for the increased number of generated outputs,
we modify the recommended temperature to 0.3 for the first leg (code-to-text) and 0.4 for the second leg (text-to-code). 

For the GPT models, we start following the advice backed by OpenAI \footnote{~\url{https://community.openai.com/t/cheat-sheet-mastering-temperature-and-top-p-in-chatgpt-api/172683}} that for code-generation tasks, LLMs should use a lower temperature when generating structured code compared to natural language generation and therefore choosing $0.3$ and $0.2$ for each leg of the translation.
We perform a temperature sweep for GPT-3.5 and GPT-4o-mini (further discussed in Section \ref{subsec:temp-sweep}, arriving at different best-performing temperatures for said models).
We decide to use top-p close to one to increase diversity but follow OpenAI guidelines to not drastically modify temperature and top-p simultaneously.
Moreover, we experimented with different other hyperparameters. 
For example, we tried different tags in the Javadoc header and values of \emph{repetition\_penalty}, 
\emph{length\_penalty}, \emph{no\_repeat\_ngram\_size}. 
Albeit we did not perform a systematic study, the 
results from other prompts and the use of these hyperparameters did not show any promising improvement in our use case.
Therefore, we have used the parameters as reported in Table~\ref{tab:hyper} and the prompts described in Section~\ref{sec:prompt-choice}.

\begin{DIFnomarkup} %
\begin{table}[t]
\centering
\caption{Hyperparameters used in RTT.}
\begin{tabular}{lrrrr}
\toprule
 Model & 
 \makecell{Number \\ of beams} & 
 \makecell{Temperature \\ first leg} & 
 \makecell{Temperature \\second leg\\} & 
 \makecell{Top-p \\ } \\ 
\midrule
PLBART & 10 & 1 & 1 & - \\
CodeT5 & 10 & 1 & 1 & - \\
TransCoder & 10 & 1 & 1 & - \\
SantaCoder & 1 & 0.3 & 0.4 & 0.95 \\
InCoder & 1 & 0.3 & 0.4 & 0.95 \\
StarCoderBase & 1 & 0.3 & 0.4 & 0.95 \\
GPT-3.5 & 1 & 1.0 & 1.0 & 0.95 \\
GPT-4o-mini & 1 & 0.6 & 0.6 & 0.95 \\
GPT-4 & 1 & 0.3 & 0.2 & 0.95 \\
\bottomrule
\end{tabular}
\label{tab:hyper}
\end{table}
\end{DIFnomarkup} %

\subsection{Prompt Choice}
\label{sec:prompt-choice}
The prompts for the instruction and cloze-style models differ due to the presence of the system message in the GPT models and the special infilling tokens for the cloze-style models.

\subsubsection{GPT models}

We do not use conversation memory in-built
for the three GPT models
and run only one forward and one backward translation step.
The \textbf{system message} is as follows: \\
\smalltt{You are an expert programmer in all programming languages.}\\
The \textbf{user prompt} for PL $\rightarrow$ NL summarization (forward translation) has been chosen in the following way: \\
\smalltt{Create a Javadoc for the Java function delimited by triple backquotes. Do not return generate the method again, return only the Javadoc. Java function: \`{}\`{}\`{}\{buggy code\}\`{}\`{}\`{}}.

For NL $\rightarrow$ PL code generation (backward translation), the \textbf{user prompt} is as follows: \\
\smalltt{Given the signature of a Java function and its Javadoc delimited by triple backquotes, generate the body of the function. Do not generate any additional methods nor repeat the Javadoc nor give any explanations. Return only the completed function without any comments.\`{}\`{}\`{}\{NL description as a comment and function signature\}\`{}\`{}\`{}}.

\subsubsection{Open-source Models}
For the rest of the models, we follow the default settings with the exception of the following modifications.
We ban the word \textit{TODO} when generating code with StarCoder and SantaCoder, since a considerable amount of generated candidate patches were left blank after the use of the word.
In addition, the tag \textit{@description} is inserted in the Javadoc header to prompt the models to generate natural language summaries after the description tag. 

The resulting prompt for the PL $\rightarrow$ NL step (forward translation) has the following form:\\[.5ex]
    \begin{minipage}{.95\columnwidth}    %
      \smalltt{/* @description <INFILL>}\\  %
      \smalltt{*/}\\                        %
      \smalltt{\{buggy code\}}.\\           %
    \end{minipage}

We prepend the NL summary with the header $<| file\ ext=.java |>$ when using InCoder for the code generation (second leg) to improve the overall results of the model by giving context. 
The prompt for NL $\rightarrow$ PL code generation (backward translation) is, therefore:\\[.5ex]
    \begin{minipage}{.95\columnwidth}    %
      \smalltt{<| file\ ext=.java |>}\\
      \smalltt{/* @description \{NL description\}}\\  %
      \smalltt{*/}\\                        %
      \smalltt{\{function signature\}}.\\            %
    \end{minipage}

\subsection{Ensuring Testability of Candidate Patches}
\label{sec:postprocess-ensure-testability}

\begin{figure}
    \centering
    \includegraphics[width=0.5\linewidth]{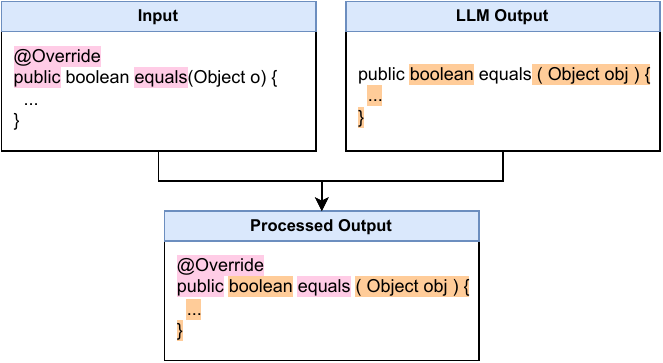}
    \caption{Post-processing step to overwrite scope and method name in the output of the LLM.}
    \label{fig:postprocessing}
    \Description[Post-processing adjusts method scope and name in language model output to match the original input.]{The figure presents a three-part diagram illustrating how post-processing corrects the scope and method name in a generated Java method. 
    The input panel shows a method definition for "equals" with the annotation "@Override", using the parameter "Object o". The language model output incorrectly changes the parameter to "Object obj", removes the annotation, and modifies the method formatting. 
    The processed output combines the correct method scope and annotation from the input with the body of the language model's generated code, resulting in a corrected and coherent output. 
    This demonstrates a step in aligning model predictions with original code structure.}
\end{figure}

When generating a candidate patch with RTT, we take two post-processing approaches according to the model used.
For models such as PLBART, CodeT5, and TransCoder, we observe that they do not retain essential contextual information such as method names, scopes, or annotation (e.g., \texttt{@Override}). 
To address this limitation, we automatically overwrite the signature of the generated candidate patch with the appropriate ones from the buggy code.
This ensures that the candidate patch is tested regardless of small errors such as a change of the function name.
This process is shown in Figure \ref{fig:postprocessing}.

In contrast, when using the rest of the models, we provide the function signature known from the original buggy example alongside the NL description generated in the first RTT leg (forward translation step).
Since function signatures are provided as context, and LLMs use conditional generation setup, it is likely that function logic will be implemented with the variables from the function signatures in the code generated by LLMs.
As a result, candidate patches from these models can be tested directly without requiring additional modifications.

Finally, we set up the RTT pipeline so that we skip a buggy example if its original code or the generated translation does not fit in the context window of the model and mark such cases
in the final results. 
However, out of the four benchmarks and eight models from our evaluation setup, it happens only for the \emph{Jsoup 15} bug and the StarCoderBase model
that we are forced to skip a buggy example due to a lack of computing resources.

\subsection{Evaluation Metrics}
We compute a total of seven common APR metrics for each candidate patch generated by the RTT pipeline to evaluate the performance of RTT with different models and assess the effectiveness of RTT for APR~\cite{zhang2023:survey}. We report the following metrics to \emph{Weights \& Biases},\footnote{~\url{https://wandb.ai}}
an online tool to analyze the models and their results:
\begin{itemize}
    \item \emph{Compilability} $\in \left\{0,1\right\}$: ability of the candidate patch to be compiled successfully.
    \item \emph{Plausibility} $\in \left\{0,1\right\}$:
    ability of the candidate patch to pass all test cases of the corresponding benchmark. 
    \item \emph{Test pass rate} $\in \left[0,100\right]$: percentage of tests passed by the candidate patch.
    \item \emph{Exact Match} $\in \left\{0,1\right\}$: binary metric to check if the candidate patch exactly matches the target solution.
    \item \emph{BiLingual Evaluation Understudy (BLEU)} $\in \left[0,1\right]$~\cite{papineni2002:bleu}: evaluates by comparing the n-grams against target solution.
    \item \emph{CodeBLEU} $\in \left[0,1\right]$~\cite{ren2020:codebleu}: extension of BLEU designed for source code. 

    The score is a weighted sum of four components: 
    \emph{(a)} BLEU, measuring n-gram overlap;
    \emph{(b)} Weighted N-Gram Match, assigning higher importance to programming-related keywords;
    \emph{(c)} Syntactic Abstract Syntax Tree (AST) Match, comparing syntactic similarity through ASTs;
    and \emph{(d)} Semantic Data-Flow Match, comparing semantic similarity through data-flow graphs.

\end{itemize}

\subsection{Manual Assessment of Correctness}
We perform manual assessment of correctness in addition to the automated test suites on the HumanEval-Java dataset.
We decide to perform this assessment only on one dataset for three reasons:
(a) HumanEval-Java has been created with the goal of reducing data leakage into the models;
(b) given the number of runs, number of models, and number of outputs per model, every new benchmark manually assessed quickly increases the amount of patches to a prohibitive amount;
(c) Defects4J contains problems that require deep knowledge about the corresponding project or a significant amount of time to verify. 
For example, a patch containing a different method being called does not always indicate an incorrect patch.
In such cases, the reviewer would need to manually inspect the additional methods, compare their logic against the ground truth, and determine if the program semantics are preserved.
The overriding of methods and different scopes makes the task of finding the executed code non-trivial. 

In the manual assessment, we compare the generated patch against the reference solution provided in the dataset.
Each reviewer examined a subset of problems focusing only on plausible patches.
For each problem and each run, we reviewed the patches in the order they were generated until we found a correct patch.
If the reviewer was unsure about the equivalence of the patch to the solution, it was collectively discussed until a consensus was agreed.

\subsection{Further Implementation Details}
\label{subsec:further-details}

\head{Model Choice:}
Forward and backward translations are performed by the same language model.
We consciously make this design choice for consistency and simplicity.
Using the same model ensures comparable performance across both legs of the translation.
We acknowledge that employing different models may offer advantages, but such configurations introduce additional complexity that can quickly increase the number of needed experiments to perform a thorough analysis.
This consideration is further discussed in Section \ref{sec:future}.
\head{Addressing Model Stochasticity:}
To account for randomness in LLMs with non-zero temperatures,
we run each experiment that uses open source models with 10 different seeds and refer to these as \emph{10 runs}. 
This helps mitigate the impact of randomness and results in a more accurate representation of RTT capabilities.
We perform only one run for each of the experiments with 
OpenAI's models, because at the moment of writing these models do not allow setting a seed.\footnote{~This has the added benefit of limiting overall experiment costs, especially for GPT-4. The approximate cost for the single run using 
GPT-4 
was \appr140USD.}

\head{Hardware:}
We run the patch generation on 3 NVIDIA V100 GPUs or 2 NVIDIA A100 GPUs, depending on model needs. 
We run the test suites for patch validation on a 32-Core AMD EPYC 7601 CPU with 2TB RAM.

\section{Results and Discussion}
\label{sec:results}
\subsection{Round-Trip Translation through PL}
\label{subsec:results-pl}
We first investigate the capabilities of LLMs to fix bugs via RTT using another programming language as the intermediate. 
For this purpose, we use the 
LLMs which are able to translate Java code into another PL:
PLBART (Java $\leftrightarrow$ C\#), 
CodeT5 (Java $\leftrightarrow$ C\#), 
and TransCoder (Java $\leftrightarrow$ C++, Java $\leftrightarrow$ Python).
Table~\ref{tab:round-trip-pl-nsolved} summarises the bug fixing performance of RTT along with the intermediate language used in RTT.
We set plausibility to 1 if at least one of the 10 generated candidate patches passes all the tests for a given buggy code sample, 
sum up plausibility over buggy code examples in a dataset, and then take an average over 10 runs with different seeds.
We refer to 
the number of unique problems with at least one plausible patch
as \emph{plausibility rate.}

\begin{DIFnomarkup} %
\begin{table}[t]
\centering
\newcommand{\bugs}[1]{(#1 bugs)} %
\caption{Average number of unique problems with at least one plausible patch ± standard deviation over 10 runs, generated with a PL as intermediate. The best results on each dataset are highlighted in \best{bold}.}
\begin{tabular}{lcccc}
\toprule
 Model & 
 \makecell{Defects4J \\ v1.2 \\ \bugs{130}} & 
 \makecell{Defects4J \\ v2.0 \\ \bugs{89}} & 
 \makecell{QuixBugs \\ \\  \bugs{40}} & 
 \makecell{Human \\ Eval-Java \\ \bugs{164}} \\ 
\midrule
PLBART (C\#) & 1.0 ± 0.0 & 0.0 ± 0.0 & 0.0 ± 0.0 & 0.0 ± 0.0 \\
CodeT5 (C\#) & 1.0 ± 0.0 & 1.2 ± 0.4 & 0.0 ± 0.0 & 1.0 ± 0.0 \\
TransCoder (C++) & 1.0 ± 0.0 & \best{3.0 ± 0.0} & 0.0 ± 0.0 & 3.0 ± 0.0 \\
TransCoder (Python) & \best{3.0 ± 0.0} & 1.0 ± 0.0 & \best{0.1 ± 0.3} & \best{5.0 ± 0.0} \\
\bottomrule
\end{tabular}
\label{tab:round-trip-pl-nsolved}
\end{table}
\end{DIFnomarkup} %

For RTT through PL, we observe that the average plausibility rate is low, 
with at most five bugs repaired on average for the HumanEval-Java dataset with 164 buggy code examples and at most three code examples repaired on average for the remaining three datasets.
PLBART only fixes a single bug across the four datasets with RTT.  
CodeT5 provides at most two plausible patches for any of the datasets.
The best performance with RTT through PL is achieved by TransCoder with Python as intermediate PL on three out of four datasets.
Moreover, TransCoder is the only model that provided a plausible patch for QuixBugs.
We observe that larger models fix more buggy examples than smaller ones, 
which is aligned with the general tendency of larger models to perform better on downstream tasks~\cite{wei2022:emergent}.

The models tend to repeat the candidate patches for a given buggy example over the 10 runs regardless of the non-zero temperature and different random seeds.
This trend, in addition to the low plausibility rates and standard deviation obtained, can indicate a potential rigidity in the conceptual mapping between languages, which may limit the model to literal translation, preventing efficient use of context to filter out noise, or bugs.
In other words, code-to-code NMT models with similar target and source languages keep the same tokens and logical bugs.
This is supported by the fact that TransCoder with Python as intermediate performs the best on three out of four datasets. 
Python and Java are less alike than C\# or C++ and Java, which motivates bigger changes when translating.

\begin{mdframed}[style=mystyle]
\head{Answer to RQ1 (RTT through PL):} 
The use of PL as an intermediate language in our approach, while yielding a very low number of plausible patches, has shed light on a few key points: \emph{(a)} the intermediate translation should differ enough from the buggy code; 
\emph{(b)} larger models produce better RTT results on APR through PL. 
\end{mdframed}

\subsection{Round-Trip Translation through NL}
We continue our experiments with RTT that uses a natural language (English) as intermediate representation.
We report the number of buggy code examples with plausible patches 
in Table~\ref{tab:results-nl-compilation}. 
We include average and standard deviation of the plausibility rate over 10 runs with different seeds, as well as the exact values observed in the union of all runs (Any Run) and their intersection (Every Run). 
Note that the models in Table~\ref{tab:results-nl-compilation} are ordered by size.     

\begin{DIFnomarkup} %
\begin{table}[t]
\newcommand{\vertical}[2]{\multirow{#1}{*}{\rotatebox{90}{#2}}}
\newcommand{\bugs}[1]{(#1 bugs)} %
\newcommand{\size}[1]{& #1}
\newcommand{\showsize}[1]{& #1}
\centering
\setlength{\tabcolsep}{7pt}
\caption{Number of unique problems with at least one plausible patch over 10 runs, with NL as intermediate. GPT-3.5 and GPT-4 are run once, and reported in the last two rows of ``Any Run''. The best results in each group are shown in \best{bold}.}
\begin{tabular}{llrrrrr}
\toprule
 & Model \size{\makecell{Model\\ size}} & 
 \makecell{Defects4J \\ v1.2 \\ \bugs{130}} & 
 \makecell{Defects4J \\ v2.0 \\ \bugs{89}} & 
 \makecell{QuixBugs \\ \\  \bugs{40}} & 
 \makecell{Human \\ Eval-Java \\ \bugs{164}} \\ 
 \midrule
\vertical{6}{Avg ± STD}

& PLBART \size{140M} & 1.0 ± 0.0 & 1.0 ± 0.0 & 2.0 ± 0.0 & 3.0 ± 0.0 \\
& CodeT5 \size{220M} & 2.0 ± 0.0 & 2.0 ± 0.0 & 1.0 ± 0.0 & 4.0 ± 0.0 \\
& SantaCoder \size{1.1B} &
\secbest{7.5 ± 1.3} &
\secbest{5.3 ± 1.7} &
\secbest{12.1 ± 1.2} &
\secbest{31.4 ± 2.0} \\
& InCoder \showsize{1.3B} & 4.0 ± 0.9 & 3.3 ± 1.5 & 4.6 ± 0.9 & 16.5 ± 1.4 \\
& InCoder \showsize{6.7B} & 4.9 ± 1.0 & 4.5 ± 0.9 & 7.8 ± 1.3 & 26.8 ± 2.4 \\
& StarCoderBase \size{15.5B} &
\best{9.2 ± 1.5} &
\best{6.3 ± 1.4} &
\best{19.8 ± 1.5} &
\best{36.9 ± 7.2}
\\
\midrule
\vertical{9}{Any Run}

& PLBART \size{140M} & 1 & 1 & 2 & 3 \\
& CodeT5 \size{220M} & 2 & 2 & 1 & 4 \\
& SantaCoder \size{1.1B} & 13 & \best{12} & 21 & 46 \\
& InCoder \showsize{1.3B} & 6 & 7 & 8 & 30 \\
& InCoder \showsize{6.7B} & 10 & 8 & 14 & 41 \\
& StarCoderBase \size{15.5B} & \best{16} & 10 & \best{24} & \best{52} \\
\cmidrule{2-7}
& GPT-3.5 \size{UNK} & 14 & 2 & 30 & 84 \\
& GPT-4o-mini \size{UNK} & 10 & 5 & \best{33} & 90 \\
& GPT-4 \size{UNK} & \best{17} & \best{7} & 27 & \best{100} \\
\midrule

\vertical{6}{Every Run}
& PLBART \size{140M} & 1 & 1 & 2 & 3 \\
& CodeT5 \size{220M} & 2 & 2 & 1 & 4 \\
& SantaCoder \size{1.1B} & 2 & 2 & 5 & \best{20} \\
& InCoder \showsize{1.3B} & 2 & 0 & 1 & 8 \\
& InCoder \showsize{6.7B} & 2 & 0 & 3 & 15 \\
& StarCoderBase \size{15.5B} & \best{6} & \best{3} & \best{16} & 16 \\
\bottomrule
\end{tabular}
\vspace{1ex}
\label{tab:results-nl-compilation}
\end{table}
\end{DIFnomarkup} %

A strong correlation is observed between the model size (excluding models with undisclosed size) and the number of compilable patches (Person's $r = 0.60$).
We also observe an even stronger correlation between the model size and the plausibility rate (Person's $r = 0.77$).
The only outlier is SantaCoder, 
which also performs comparably to larger models on other tasks in related work~\cite{allal2023:santacoder}.
The growth of the average plausibility rate from SantaCoder (1.1B) to StarCoderBase (15.5B) is less pronounced.
This result can be affected by a larger proportion of Java code within SantaCoder training data compared to StarCoderBase training set.
In addition, the majority of models with more than 1B parameters fix at least one bug with RTT through NL that is not repaired by RTT through NL with other models. 
Note that for a fixed model, bug repair performance differs based on the dataset. 
Thus, the proportion of the buggy input examples with plausible patches is lower for more complex datasets (Defects4J variants) and larger for simpler tasks (QuixBugs and HumanEval-Java).

The low standard deviation of the plausibility rate indicates that most models tend to repair a 
similar number of bugs in every run.
However, the number of fixed bugs on \emph{Any Run} is two to three times higher on average than the number of repaired bugs in \emph{Every Run}.
For example, StarCoderBase fixes 16 Defects4J v1.2 buggy input examples in the aggregation of 10 runs and 
only 6 same bugs in every run.
This indicates that RTT is capable of repairing diverse bugs.

\begin{figure}[t]
\centering
    \includegraphics[width=0.8\textwidth]{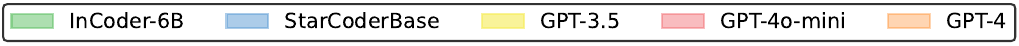}
\begin{subfigure}{0.32\textwidth}
    \includegraphics[width=\textwidth]{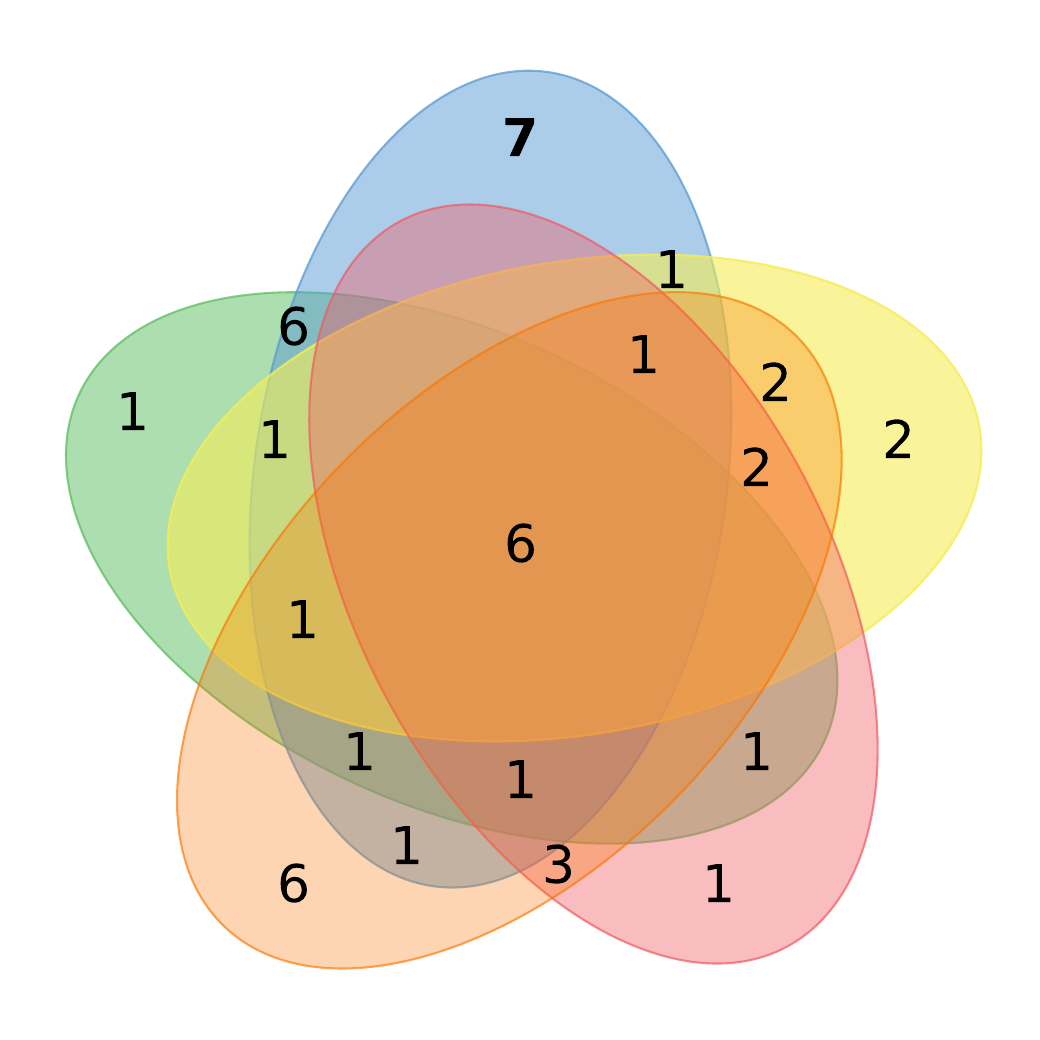}
    \caption{Defects4J}
    \label{fig:venn-solved-defects4j}
\end{subfigure}
\begin{subfigure}{0.32\textwidth}
    \includegraphics[width=\textwidth]{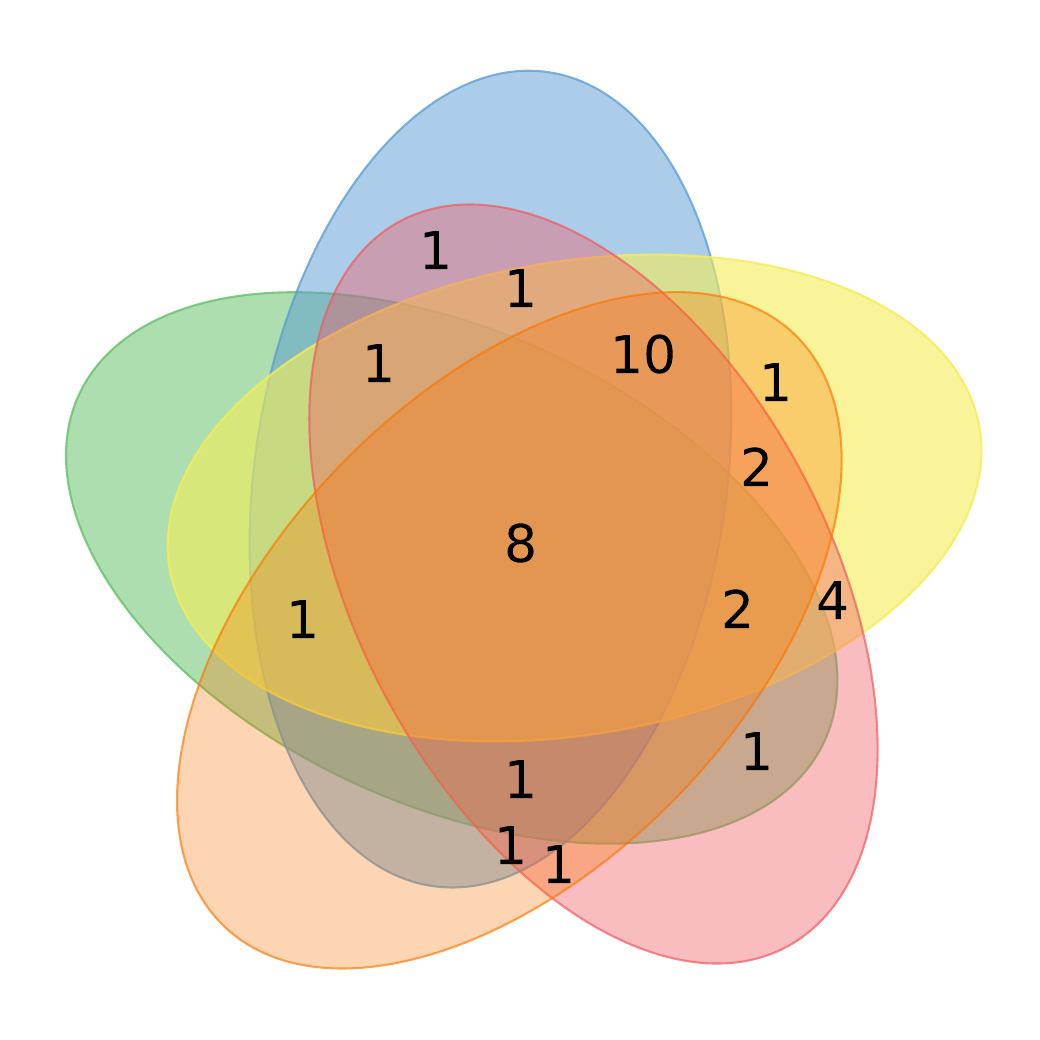}
    \caption{QuixBugs}
    \label{fig:venn-solved-quixbugs}
\end{subfigure}
\begin{subfigure}{0.32\textwidth}
    \includegraphics[width=\textwidth]{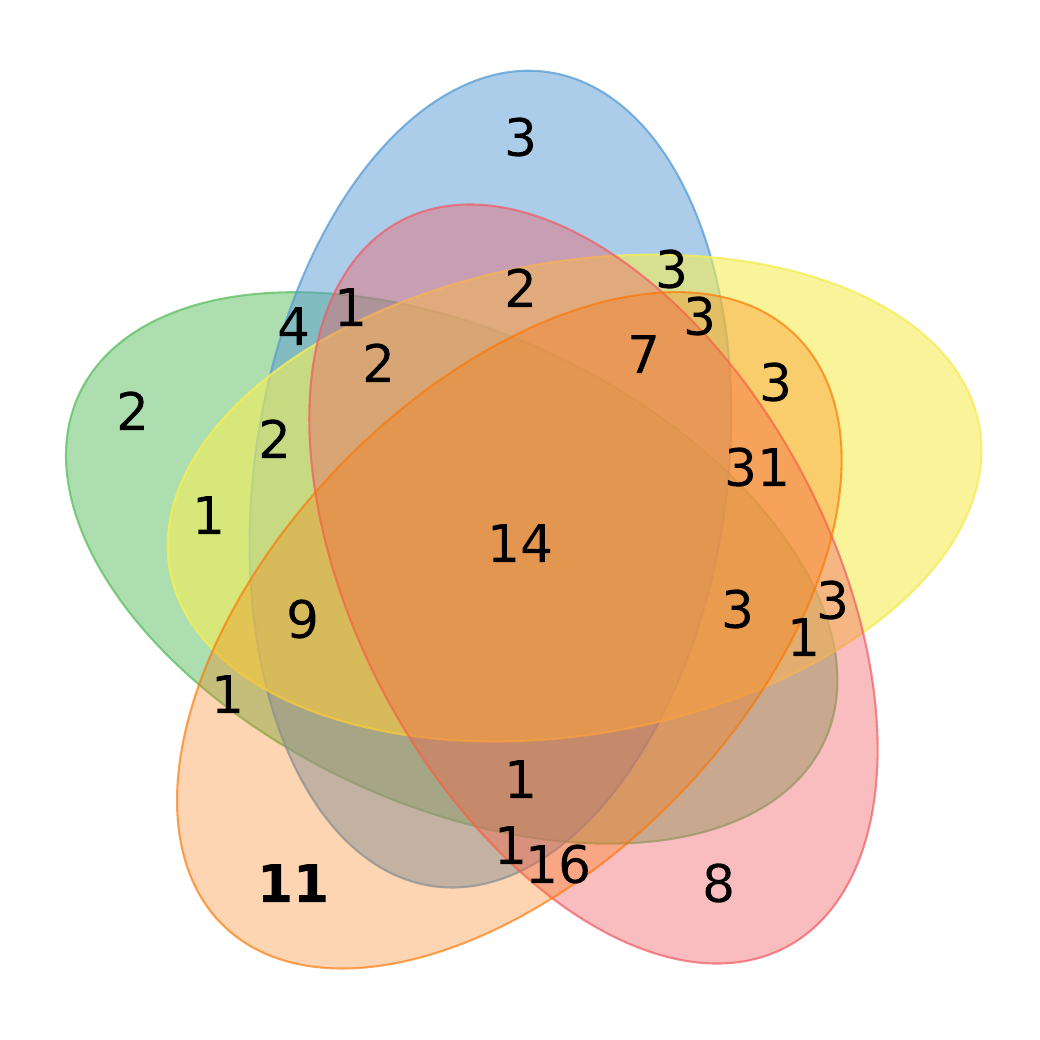}
    \caption{HumanEval-Java}
    \label{fig:venn-solved-humaneval}
\end{subfigure}
\caption{
Number of unique bugs fixed in various datasets by RTT through NL with a fixed language model. The largest number for each dataset is highlighted in \textbf{bold}.}
\label{fig:venn-solved-all}
\Description[Unique bugs fixed by language models across three datasets using round-trip translation through natural language.]{
The figure presents three Venn diagrams comparing the number of unique bugs fixed by five different language models—InCoder-6B, StarCoderBase, GPT-3.5, GPT-4o-Mini, and GPT-4—across the Defects4J, QuixBugs, and HumanEval-Java datasets. 
Each diagram shows the overlap and uniqueness in bugs fixed among models, with the largest individual count per dataset displayed in bold. 
In the Defects4J dataset, StarCoderBase fixed the most unique bugs (7). 
In QuixBugs, there is none of the models solve a unique problem, and in HumanEval-Java, GPT-4 fixed the most (11). 
The diagrams reveal both overlapping capabilities and distinct strengths among the models in repairing bugs across different benchmarks.}
\end{figure}

The aggregated metrics over 10 runs bring additional perspectives, such as a comparison of unique bugs fixed by RTT and those in related work.
Although the number of repaired bugs tends to vary between the models, Figure \ref{fig:venn-solved-all} indicates that in our approach, most models tend to solve at least one unique problem.
Therefore, the kind of problems solved by RTT are not only size-dependent, but also model-dependent.

\subsubsection{Comparison with previous work}
\label{sec:comparison-with-previous-work}

\begin{DIFnomarkup} %
\begin{table}[t]
\newcommand{\vertical}[2]{\multirow{#1}{*}{\rotatebox{90}{#2}}}
\newcommand{\bugs}[1]{(#1 bugs)} %
\newcommand{\size}[1]{& #1}
\newcommand{\showsize}[1]{& #1}
\centering
\caption{Number of unique problems with at least one plausible patch from previous work (\citet{jiang2023:impact}).}
\begin{tabular}{llrrrrr}
\toprule
 & Model \size{\makecell{Model\\ size}} & 
 \makecell{Defects4J \\ v1.2 \\ \bugs{130}} & 
 \makecell{Defects4J \\ v2.0 \\ \bugs{89}} & 
 \makecell{QuixBugs \\ \\  \bugs{40}} & 
 \makecell{Human \\ Eval-Java \\ \bugs{164}} \\ 
 \midrule
\vertical{5}{Base Models}
        & PLBART \size{140M} & \best{25} & \best{25} & 13 & 40 \\
        & CodeT5 \size{220M} & 3 & 7 & 0 & 5 \\
        & InCoder \showsize{1.3B} & 13 & 19 & \best{18} & 40 \\
        & CodeGen \size{2B} & 14 & 6 & 17 & 50 \\
        & InCoder \showsize{6.7B} & 20 & 20 & \best{18} & \best{59} \\
        \midrule
        \vertical{5}{\makecell{Fine-Tuned \\ Models}}
        & PLBART \size{140M} & 33 & 24 & 15 & 36 \\
        & CodeT5 \size{220M} & 33 & 25 & 17 & 54 \\
        & InCoder \showsize{1.3B} & 43 & \best{38} & 20 & 64 \\
        & CodeGen \size{2B} & 38 & 36 & 20 & 53 \\
        & InCoder \showsize{6.7B} & \best{56} & \best{38} & \best{24} & \best{70} \\
        \midrule
        \vertical{4}{\makecell{DL-based\\ APR}}
        & CURE \size{N/A} & 6 & 6 & 5 & 18 \\
        & Reward \size{N/A} & 20 & 8 & 7 & \best{22} \\
        & Recoder \size{N/A} & \best{24} & 11 & 6 & 11 \\
        & KNOD \showsize{N/A} & 20 & \best{13} & \best{10} & 18 \\
\bottomrule
\end{tabular}
\vspace{1ex}
\label{tab:results-from-previous-work}
\end{table}
\end{DIFnomarkup} %
 
To position RTT in the broader APR landscape, we compare the results to those reported in prior studies.
For easier comparison we include 
Table~\ref{tab:results-from-previous-work} which compiles the results from models of varying sizes, base and fine-tuned for APR, as well as some DL-based APR techniques from \citet{jiang2023:impact}.
Additionally, Table~\ref{tab:comparison-old-work-finetuned} compares the results of our RTT approach with 
the same fine-tuned models to find the intersection of problems solved.
We made this comparison with problems that have at least one plausible patch based on the results reported in their repository since the authors have not released their manual assessment.
The majority of models used in RTT repair at least one buggy example not repaired by the same models fine-tuned for NMT-type of APR methods.

We also compare 
our
top-performing model 
for HumanEval-Java
, GPT-4, against the 10 LLMs studied in the work of \citet{jiang2023:impact}, fine-tuned and non-finetuned for the APR task, in Figure~\ref{fig:venn-solved-gpt4}.
Not only is GPT-4 able to generate more plausible patches than any of the tested models by Jiang et al.~\cite{jiang2023:impact}, but 30 out of the 100 bugs are only fixed by RTT through NL and not repaired by any of the tested models without RTT (see Figure~\ref{fig:venn-solved-gpt4}).
This comparison highlights that RTT generates plausible patches for the bugs that common APR approaches 
have not fixed and emphasizes the added value of RTT in the APR landscape.

Studies show that generating more candidate patches can result in higher repair performance on the datasets used in our evaluation~\cite{xia2023:keep, xia2023:automated}.
The authors generate 200 patches per model, compared to 
10
patches in our case, which may be the reason why they fix all bugs in QuixBugs with GPT-3.5~\cite{xia2023:keep} or get 26 and 29 plausible patches with InCoder 1.3B and 6.7B, respectively~\cite{xia2023:automated}.
However, we could not find enough details or replication packages for those studies and do not include them in Table~\ref{tab:comparison-old-work-finetuned}.
The latter work also shows that plausible patches have, on average, lower entropy than non-plausible ones.
In our experiments, we obtain an almost uniform distribution of the number of plausible patches depending on their position, i.e., over how far on the list of 
candidate patches the plausible ones occur. 
This suggests that results can be improved by generating more candidate patches with RTT, at higher resource usage.

\begin{DIFnomarkup} %
\begin{figure}[t]
\begin{minipage}{0.55\textwidth}
\captionsetup{width=0.98\textwidth}
\vspace{5.6ex}
    \centering
    \small
    \newcommand{\bugs}[1]{(#1 bugs)} %
    \newcommand{\triple}[3]{#1 / #2 / #3}
    \setlength{\tabcolsep}{5pt}
    \captionof{table}{Number of unique problems with at least one plausible patch, shown as \triple{P}{O}{N}, with P in previous work, O in our work using RTT through NL (\textit{Any Run}), and N only in our work, not in previous.}
    \begin{tabular}{lrrrr}
    \toprule
     Model  & 
     \makecell{Defects4J \\ v1.2 \\ \bugs{130}} & 
     \makecell{Defects4J \\ v2.0 \\ \bugs{89}} & 
     \makecell{QuixBugs \\ \\  \bugs{40}} & 
     \makecell{Human \\ Eval-Java \\ \bugs{164}} \\ 
     \midrule
    PLBART         & \triple{33}{\enspace1}{0}   & \triple{24}{1}{1}   & \triple{15}{\enspace2}{1} & \triple{36}{\enspace3}{\enspace0}            \\
    CodeT5         & \triple{33}{\enspace2}{1}   & \triple{25}{2}{1}   & \triple{17}{\enspace1}{1} & \triple{54}{\enspace4}{\enspace0}            \\
    InCoder (1.3B) & \triple{43}{\enspace6}{3}   & \triple{38}{7}{2}  & \triple{20}{\enspace8}{4} & \triple{64}{30}{11}          \\
    InCoder (6.7B) & \triple{56}{10}{3}  & \triple{38}{8}{1}  & \triple{24}{14}{4} & \triple{70}{41}{16}          \\
    \bottomrule
    \end{tabular}
    \vspace{4ex}
      \label{tab:comparison-old-work-finetuned}
    \end{minipage}
\hfill
  \begin{minipage}{0.44\textwidth}
  \captionsetup{width=0.8\textwidth}  
    \centering
    \includegraphics[width=0.78\textwidth, trim={0.4cm 1.9cm 0.4cm 0},clip]{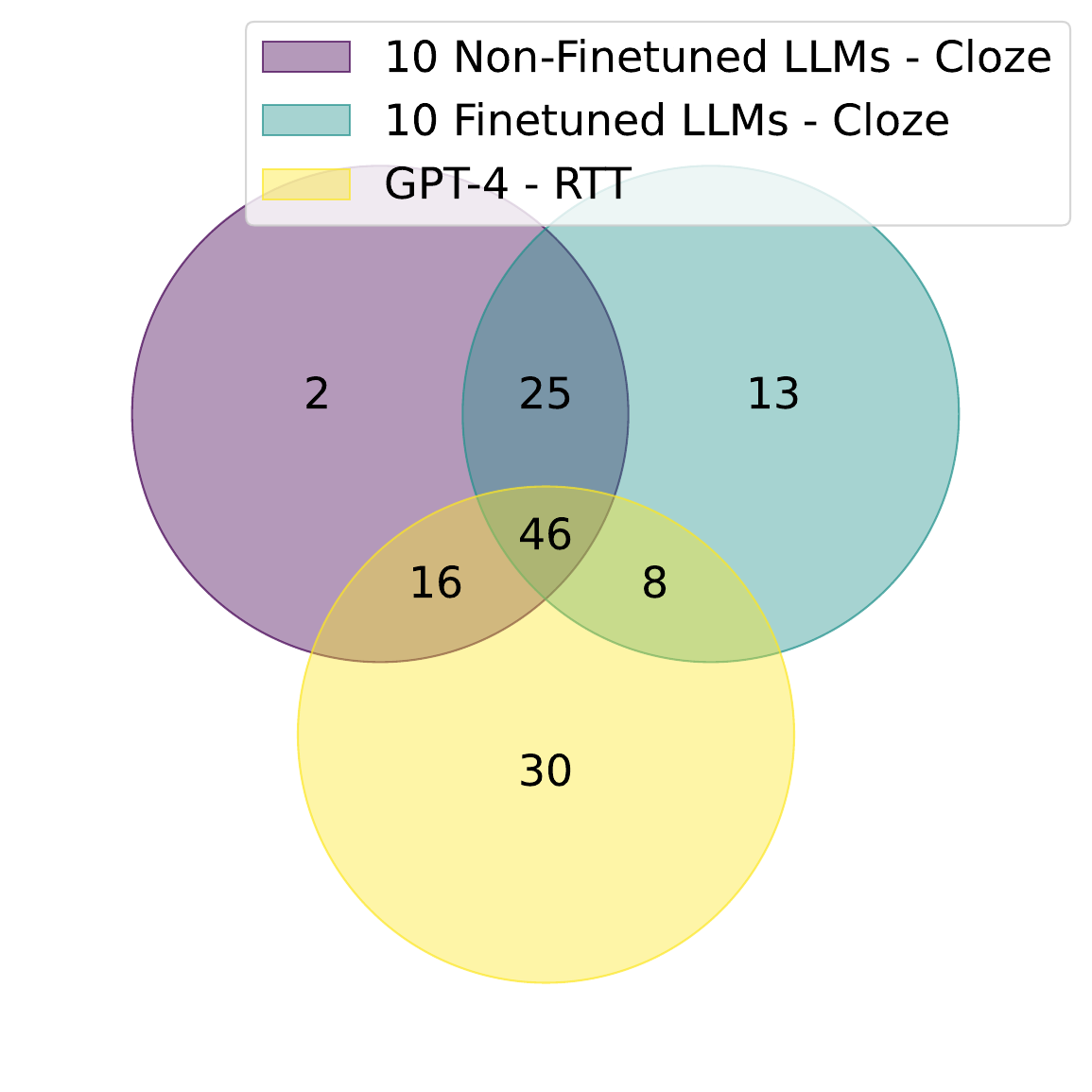}
    \captionof{figure}{Comparison of GPT-4 on the number of HumanEval-Java problems with plausible patches.}
    \label{fig:venn-solved-gpt4}
    \Description[Venn diagram showing overlap of correct solutions among GPT-4 and two groups of language models on HumanEval-Java.]{The diagram illustrates the number of problems correctly solved by three model groups: 10 non-finetuned language models using the cloze format, 10 finetuned language models using the cloze format, and GPT-4 using Round-Trip-Translation (RTT).
    GPT-4 uniquely solved 30 problems, and 46 were solved by all three groups. 
    The non-finetuned group solved 2 problems uniquely, while the finetuned group solved 13 uniquely. 
    Overlaps between pairs include 25 problems solved by both non-finetuned and finetuned models, 16 by GPT-4 and non-finetuned models, and 8 by GPT-4 and finetuned models. 
    The total number of evaluated problems is represented by the union of all sets.}
  \end{minipage}
\end{figure}
\end{DIFnomarkup}

\subsubsection{HumanEval-Java Manual Correctness Assessment}
We conducted a manual assessment of the correctness of RTT-generated patches for the HumanEval-Java benchmark.
Table~\ref{tab:correct-fixes-comparison} reports the number of manually verified correct fixes for each model in our approach, as well as in previous work~\cite{jiang2023:impact}.
GPT-4 maintains its position as the top performer for the benchmark.
On the other hand, the model with most number of overfitting patches is GPT-3.5 with a total of 17.
Although this model tied with GPT-4o-mini in the number of plausible patches, the manual analysis points out the superiority of the later model.
For the rest of the models, we see a slight decrease in the average and standard deviation of problems solved.
We further appreciate a slightly more substantial decrease in the number of problems solved in Any Run.

To promote transparency and reproducibility, our manual assessment of over 5,000 patches is released alongside our replication package.
We hope to enable other researchers to examine and/or extend our work.

\begin{DIFnomarkup} %
\begin{table}[t]
\newcommand{\vertical}[2]{\multirow{#1}{*}{\rotatebox{90}{#2}}}
\newcommand{\bugs}[1]{(#1 bugs)}
\newcommand{\size}[1]{& #1}
\newcommand{\showsize}[1]{& #1}
\centering
\caption{Comparison of number 
of unique problems with at least one correct patch (manually checked) of RTT and previous work. The best results in each group are shown in \best{bold}.}
\begin{minipage}{0.49\linewidth} %
    \centering %
    \begin{tabular}{llrrr}
        \toprule
                  & Model & \makecell{Model\\ size} & \multicolumn{2}{c}{\makecell{Human Eval-Java \\ \bugs{164}}} \\ 
         \cmidrule(lr){4-5}
         &  &  & Avg ± STD & Any Run \\ 
        \midrule
        \vertical{9}{RTT Approach}
        & PLBART \size{140M} & 3.0 ± 0.0 & 3 \\
        & CodeT5 \size{220M} & 4.0 ± 0.0 & 4 \\
        & SantaCoder \size{1.1B} & 29.4 ± 1.4 & 45 \\
        & InCoder \showsize{1.3B} & 16.2 ± 1.4 & 27 \\
        & InCoder \showsize{6.7B} & 26.0 ± 2.7 & 39 \\
        & StarCoderBase \size{15.5B} & \best{35.8 ± 7.3} & \best{51} \\
        \cmidrule{2-5}
        & GPT-3.5 \size{UNK} & -- & 80 \\
        & GPT-4o-mini \size{UNK} & -- & 88 \\
        & GPT-4 \size{UNK} & -- & \best{97} \\

        \bottomrule
    \end{tabular}
\end{minipage}%
\begin{minipage}{0.45\linewidth} %
    \centering
    \begin{tabular}{llrr}
        \toprule

         & Model & \makecell{Model\\ size} & \makecell{Human \\ Eval-Java \\ \bugs{164}} \\ 
        \midrule
        \vertical{5}{Base Models}
        & PLBART \size{140M} & 39 \\
        & CodeT5 \size{220M} & 5 \\
        & CodeGen \size{2B} & 49 \\
        & InCoder \showsize{1.3B} & 40 \\
        & InCoder \showsize{6.7B} & \best{59} \\
        \midrule
        \vertical{5}{\makecell{Fine-Tuned \\ Models}}
        & PLBART \size{140M} & 41 \\
        & CodeT5 \size{220M} & 54 \\
        & CodeGen \size{2B} & 53 \\
        & InCoder \showsize{1.3B} & 64 \\
        & InCoder \showsize{6.7B} & \best{70} \\
        \midrule
        \vertical{4}{\makecell{DL-based\\ APR}}
        & CURE \size{N/A} & 18 \\
        & Reward \size{N/A} & \best{22} \\
        & Recoder \size{N/A} & 11 \\
        & KNOD \showsize{N/A} & 18 \\
        \bottomrule
    \end{tabular}
\end{minipage}
\label{tab:correct-fixes-comparison}
\end{table}
\end{DIFnomarkup} %

\begin{mdframed}[style=mystyle]
\head{Answer to RQ2 (RTT through NL):}
We observe that 
\emph{(a)}~the trend that larger models obtain better results for bug fixing with RTT;
\emph{(b)}~although standard deviation of plausibility rate is low, the number of bugs repaired in the union of runs is 2-3 times higher on average than the rate for every run, which highlights that RTT is capable of repairing diverse bugs;
\emph{(c)}~RTT through NL is able to repair bugs not repaired by the same models fine-tuned for APR.
\end{mdframed} 

\subsection{Temperature Sensitivity Analysis}
\label{subsec:temp-sweep}
Temperature plays an important role in controlling the diversity of outputs from large language models. 
While lower temperatures increase determinism (although do not enforce it~\cite{ouyang2025:empirical}), higher temperatures promote creative and varied responses. 
In our main experiments, we choose temperatures recommended by the authors (Section~\ref{section:hyperparemeters}) for simplicity. 
However, we further analyze its effect on the RTT approach by performing a sweep on the models with the best results, GPT-3.5 and GPT-4o-mini, with six temperatures ranging from $T=0.0$ (highly deterministic) to $T=1.0$ (highly stochastic) to assess the effects of temperature in our proposed technique.
Although GPT-4 achieves the second best results, it is significantly more expensive than any other model and therefore not considered as a candidate for this experiment.
We restricted this sweep to two models since running a wide-range hyperparameter exploration for all models would be 
prohibitively expensive in terms of computation.
Furthermore, narrowing this sweep helps reduce the computational and energy footprint of our experiments, aligning with sustainable research practices.

Table~\ref{tab:temperature-sweep} summarizes the number of plausible patches produced by these two models under 
five temperatures.
For GPT-3.5, the performance generally improves with higher temperatures, reaching its peak at $T=1.0$ with 130 total plausible patches. 
On the other hand, GPT-4o-mini reaches its best results (138 total) at $T=0.6$, indicating it benefits from some stochasticity. 
This difference indicates that the optimal temperature can differ considerably across model versions or architectures.
Interestingly, for GPT-3.5, there is a monotonic improvement in the total number of repaired bugs as we increase temperature.
However, $T=1.0$ achieves the best results only in half of the benchmarks (Defects4J v1.2 and HumanEval-Java), underscoring that there is no best temperature in every scenario.
In a similar trend, GPT-4o-mini shows an optimal temperature around $T=0.6$, but $T=0.4$ still achieves a better performance on the QuixBugs benchmark.

These observations tie back to the potential rigidity discussed in Section~\ref{subsec:results-pl}, where code-to-code machine translation with closely related languages repeatedly produced the same candidate patches.
By raising the temperature, we introduce more noise into the generation process, which can help the model deviate from literal translations and therefore avoid translating the buggy logic. 
In other words, while low-temperature decoding preserves the same structure and tokens (potentially retaining the bug), higher-temperature decoding results in more diverse translations. 
This diversity is more likely to break away from any literal mapping that preserves the same error. 

\begin{mdframed}[style=mystyle]
\head{Answer to RQ3 (Temperature Senstitivity):}
We note that 
(a) temperature affects RTT repair performance, and tuning it can result in meaningful gains; (b) model-specific experimentation is crucial, as we see GPT-3.5 and GPT-4o-mini exhibit distinct optimal temperature settings. 
\end{mdframed} 

\begin{DIFnomarkup} %
\begin{table}[t]
    \centering
    \caption{Number of unique problems with at least one plausible patch for each benchmark at different temperatures (T). The best temperature for each model is shown in \textbf{bold}.}
    \begin{tabular}{lrrrrrrrrrrrr}
        \toprule
        \multirow{2}{*}{Benchmark} & \multicolumn{6}{c}{GPT-3.5} & \multicolumn{6}{c}{GPT-4o-mini} \\
        \cmidrule(r){2-7} \cmidrule(l){8-13}
         & T=0.0 & T=0.2 & T=0.4 & T=0.6 & T=0.8 & \textbf{T=1.0} & T=0.0 & T=0.2 & T=0.4 & \textbf{T=0.6} & T=0.8 & T=1.0 \\
        \midrule
        Defects4J v1.2      & 10 & 9  & 9  & 11 & 13 & \textbf{14} & 6  & 9  & 9  & 10 & \textbf{11} & \textbf{11} \\
        Defects4J v2.0      & 2  & 4  & \textbf{5}  & 3  & \textbf{5}  & 2  & 3  & 3  & 4  & \textbf{5}  & 3  & 2  \\
        QuixBugs            & 26 & 30 & 30 & 29 & \textbf{31} & 30 & 28 & 32 & \textbf{34} & 33 & 31 & 33 \\
        HE-Java             & 39 & 58 & 68 & 79 & 74 & \textbf{84} & 57 & 77 & 78 & \textbf{90} & 84 & \textbf{90} \\
        \midrule
        \textbf{Total}  & 77 & 101 & 112 & 122 & 123 & \textbf{130} & 94 & 121 & 125 & \textbf{138} & 129 & 136 \\
        \bottomrule
    \end{tabular}
    \label{tab:temperature-sweep}
\end{table}
\end{DIFnomarkup} %

\subsection{Quantitative
Analysis of Generated Candidate Patches}
Through a close inspection of candidate patches generated, 
we are able to gain more insights into the quality of RTT-generated patches.
To do this, we investigate compilability, test pass rates, CodeBLEU and other patch candidates' characteristics.

\subsubsection{Compilability} 
\begin{figure}[b]
     \centering
     \begin{subfigure}[b]{0.4\textwidth}%
         \centering
         \includegraphics[width=\textwidth]{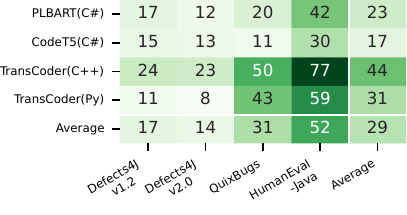}
         \caption{with PL as intermediate language\\[15pt]}
         \label{fig:heatmap-compilable-pl}
     \end{subfigure}
     \hfill
     \begin{subfigure}[b]{0.4\textwidth}%
         \centering
         \includegraphics[width=\textwidth]{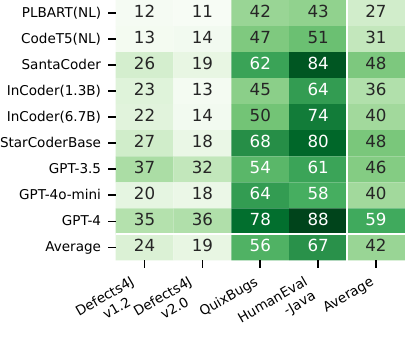}
         \caption{with NL as intermediate language.}
         \label{fig:heatmap-compilable-nl}
     \end{subfigure}
        \caption{Percentage of compilable candidate patches generated in 10 runs, where applicable, and at 10 attempts for each buggy example.}
        \label{fig:heatmap-compilable}
    \Description[Comparison of compilable patch generation rates using programming and natural languages as intermediate representations.]{The figure presents two heatmaps comparing the percentage of compilable candidate patches generated across four benchmark datasets—Defects4J  v1.2, Defects4J v2.0, QuixBugs, and HumanEval-Java—when using programming languages (left) and natural languages (right) as intermediate representations. 
    Each cell shows the percentage averaged over 10 attempts, with models listed along the vertical axis. 
    Natural language models generally outperform programming language models in average compilability across all benchmarks, especially for QuixBugs and HumanEval-Java, indicating a performance advantage when using natural language as an intermediate representation.}
\end{figure}

We explore the average ratio of compilable patches out of all candidate patches generated over 10 runs and present the results in Figures~\ref{fig:heatmap-compilable-pl} and~\ref{fig:heatmap-compilable-nl} for RTT through PL and NL, correspondingly.
In general, RTT through PL generates less compilable patches than RTT through NL. 
TransCoder helps RTT through PL generate a high number of compilable patches, which we associate with its rigorous denoising pre-training objectives. 

For RTT through NL, the trend of obtaining better compilability ratios with larger models holds.
The previously discussed pattern, where SantaCoder obtains better results than InCoder and slightly worse than StarCoderBase, is observed here, too.
Overall,
compilability on the datasets with challenging contexts (Defects4J's) is lower than on simpler tasks. 
Although compilability rates vary across the models and datasets, the majority (80-96\%) of candidate patches generated by RTT have low test pass rate, with only 0 to 10\% actually passing the test-suite (further discussed in Section \ref{subsect:test-pass-rate}).

The trends observed for plausibility rates are also present for the ratio of compilable candidate patches.
Similarly to plausibility rates, compilability percentage is higher for experiments with NL than for RTT through PL. 
On average, the percentage of compilable patches out of all generated candidate patches ranges from 8\% (with TransCoder through Python, see Figure~\ref{fig:heatmap-compilable-pl}) to 77\% (with TransCoder through C++)
for RTT through PL and from 
11\% (with PLBART, see Figure~\ref{fig:heatmap-compilable-nl}) to 88\% (with GPT-4)
for RTT through NL. 
Moreover, RTT generates a higher proportion of compilable candidate patches on average for datasets with simpler tasks (QuixBugs, HumanEval-Java) than for datasets with more complex contexts and bugs (Defects4J variants).
GPT-4, StarCoderBase and SantaCoder
are the leading models in terms of the average compilability for the RTT through NL
.
However, in the RTT through PL, the best average results are obtained with TransCoder (Java $\leftrightarrow$ C++), unlike for plausibility rate which was the best for TransCoder (Java $\leftrightarrow$ Python).

\subsubsection{Test Pass Rate}
\label{subsect:test-pass-rate}
To further explore the properties of RTT-generated candidate patches, we analyze the test pass rate of the non-plausible patches.
We aim to explore whether compilable patches pass more or less tests, i.e., whether non-plausible patches need a lot of further updates or not.

\begin{figure}
     \centering
     \begin{subfigure}[b]{0.49\textwidth}%
        \centering
        \includegraphics[width=\textwidth]{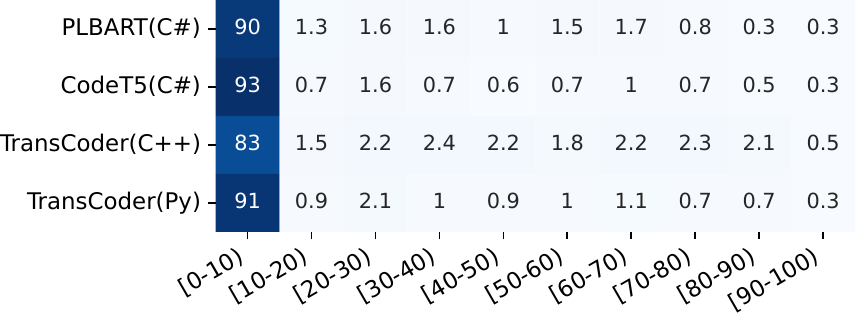}
        \caption{with PL as intermediate.\\[15pt]}
        \label{fig:heatmap-test-pass-rate-pl}
    \end{subfigure}
    \hfill
    \begin{subfigure}[b]{0.49\textwidth}%
        \centering
        \includegraphics[width=\textwidth]{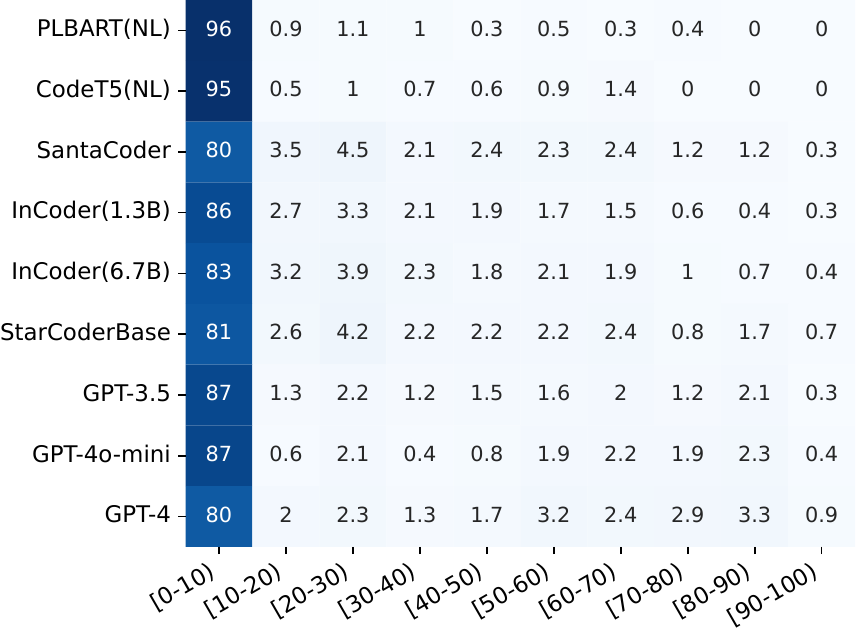}
        \caption{with NL as intermediate.}
        \label{fig:heatmap-test-pass-rate-nl}
    \end{subfigure}
        \caption{Percentage of candidate patches in the different test pass rate ranges. We explore what ratio of candidate patches generated by a specific models for any of the datasets pass from A\%~(incl.) to B\%~(excl.) tests and report percentage over all generated candidate patches. For example, 96\% of candidate patches generated with PLBART with NL as intermediate pass between 0\%~(incl.) and 10\%~(excl.) of tests.}
        \label{fig:heatmap-test-pass-rate}
    \Description[Test pass rate distribution across candidate patches using programming language and natural language intermediates.]{This figure shows two heatmaps comparing the distribution of candidate patches across test pass rate intervals when using either a programming language or natural language as an intermediate representation. 
    The left heatmap (a) displays results for PLBART, CodeT5, and TransCoder models with programming language as the intermediate. 
    The right heatmap (b) shows results for models such as PLBART, CodeT5, SantaCoder, and several GPT variants using natural language as the intermediate. 
    Each cell represents the percentage of patches whose test pass rate falls within a specific range, with darker blue indicating higher percentages. 
    In both heatmaps, the leftmost column shows the percentage of candidate patches falling in the 0 to 10 percent test pass range, which is the highest for most models. The rightmost columns, representing higher test pass rates, generally have lower values, indicating fewer high-quality patches. Overall, models using natural language intermediates show more dispersion across the test pass ranges compared to those using programming language intermediates.}
\end{figure}

The test pass rate results are presented in 
Figures~\ref{fig:heatmap-test-pass-rate-pl} and 
\ref{fig:heatmap-test-pass-rate-nl} 
for RTT through PL and NL, respectively.
We take a union of all non-plausible candidate patches over four benchmarks and all runs for each model and report the percentage of candidate patches in these unions that fall into test pass rate ranges from 0--10\%, 10--20\% and so on, including the beginning and excluding the end of the intervals. 
The vast majority of such a union of RTT-generated candidate patches pass 0 to 10\% of test cases both for PL and NL as intermediate. 
This result indicates that non-plausible patches require substantial fixing updates to repair the bugs in the chosen benchmarks.
One avenue for future work is to experiment with more iterations in the RTT (further discussed in Section \ref{sec:future}), or update the model prompts or descriptions with bug summaries or results of not passing test cases, similarly to Grishina et al.~\cite{grishina2025:fully}.

\subsubsection{Characteristics of RTT-generated candidate patches}
\begin{figure}
  \centering
  \includegraphics[width=0.5\columnwidth]{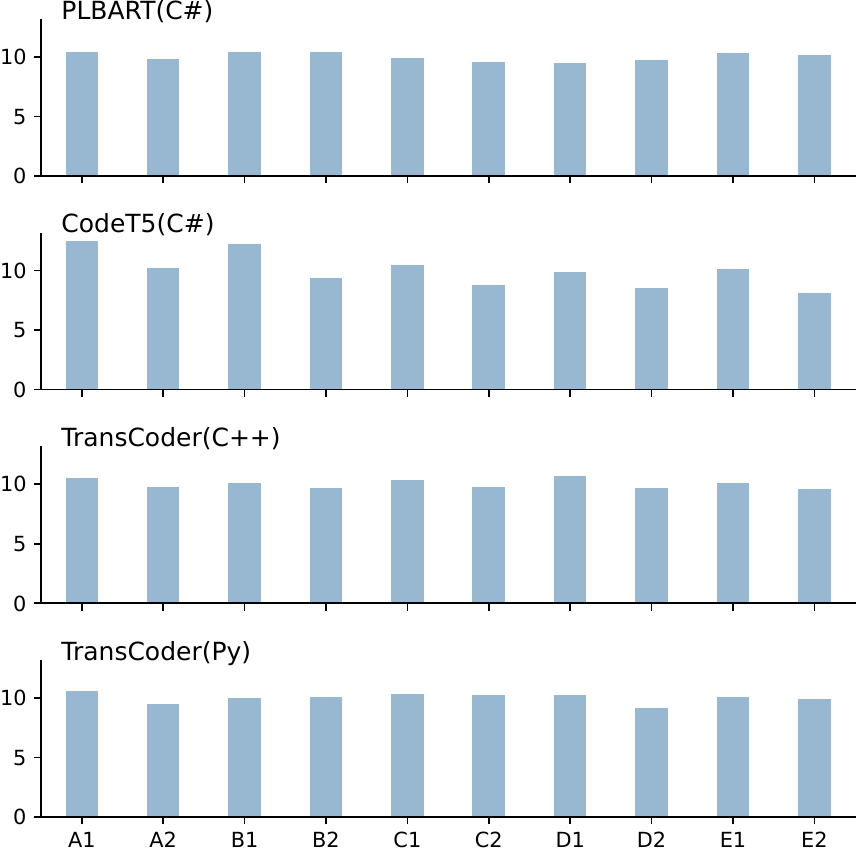}
  \caption{Percentage of compilable candidate patches in the 10 positions with PL as intermediate. The percentage is calculated over all patches in four benchmark datasets generated with RTT using a fixed model}
  \label{fig:pointplot-compilable-ranking}
  \Description[Bar chart comparing compilable patch percentages for four models across ten dataset subsets.]{The figure presents the percentage of compilable candidate patches among the top ten generated patches for each of four benchmark datasets, grouped by programming language and model.
  Four bar plots correspond to the models PLBART with C#, CodeT5 with C#, TransCoder with C++, and TransCoder with Python. 
  Each plot includes ten bars labeled from A1 to E2, representing specific dataset configurations.}
\end{figure}

To reiterate, we generate five intermediate translations in the first RTT leg, for example, in English language,
and two final translations from each of the intermediates.
We denote the first five intermediate translations as A, B, C, D, and E.
We enumerate backward translations obtained from each of the two forward translations as A1--A2, B1--B2, ..., E1--E2. 
The ratio of compilable patches out of a union over all runs and datasets of all generated patches with a fixed model is presented in Figure~\ref{fig:pointplot-compilable-ranking} for RTT through PL.  
 
With RTT through PL,
the number of compilable patches is decreasing from the first to the last candidate patch generated from a fixed intermediate translation (fixed letter).
The percentage of compilable patches in the first position (A1, ..., E1) is always higher or equal to the compilability rate at the second position for CodeT5 (C\#), and TransCoder (C++) and TransCoder (Python). The only exception is PLBART (C\#) where the compilability rate at D2 is higher than at D1.
The number of plausible patches obtained with RTT through PL is below five on average, as mentioned 
previously. 
Thus, the trend between plausibility and the position Ax,..., Ex is not observable from such low average plausibility rates.

For RTT through NL, we do not observe any trend in terms of how frequently first (A1, B1, ..., E1) or second (A2, ..., E2) patches are plausible or compilable.
Remarkably, RTT through NL and RTT through PL with TransCoder (Python) have higher plausibility rates than RTT with other models and through other PLs, as shown 
previously.
This observation points back at the discussion of rigidity of code-to-code translation models in the RTT setting: They keep same variable names, other tokens and logical bugs in place.
By contrast, NL models and code-to-code models with a PL that differs enough from the original PL show better results in  RTT for bug fixing.
They abstract and change the input buggy code enough to obtain a different representation that can in the next step lead to a bug fix.
The uniform estimated distribution of compilable and plausible patches over the position at which they are generated also supports the argument that sampling more candidate patches from LLMs in the RTT pipeline can improve the bug fixing scores.

\subsubsection{CodeBLEU}
We calculate average CodeBLEU values for input buggy examples and candidate patches generated by RTT with each fixed model over all runs and datasets and show the frequency of observed values in Figure~\ref{fig:histplot-codebleu}.
The metric values are scaled to $[0; 100],$ with highest values being the best.
The majority of buggy examples have high CodeBLEU scores, which indicates that target bug fixes are very similar to original buggy code.
High CodeBLEU for the majority of buggy examples is also explained by the type of bugs: We only consider single-hunk bugs.

In contrast,
the majority of candidate patches obtained with RTT through PL have CodeBLEU between 40 and 60, with the outlier value of ca.~26 frequently observed among candidate patches for Defects4J variants.
The most frequently observed average CodeBLEU values for RTT through NL are between 20 and 50, with a similar outlier value ca.~26 and an additional outlier of zero CodeBLEU noticeable for a number of candidate patches for Defects4J versions.
The frequency of higher CodeBLEU values increases with larger model sizes.
CodeBLEU values are considerably lower than 100 for the vast majority of RTT-generated candidate patches.
However, one can observe the trend of regressing towards the mean type of candidate patches with similar CodeBLEU calculated between targets and RTT-generated patches. 

Furthermore, although larger models shift the distribution of generated patches to higher CodeBLEU scores, 
they still remain far below the scores of the input distribution.
This indicates that substantially bigger models, although increasing plausibility rates, still produce patches significantly different from the proposed solutions.
We further expand on this observation by noting a weak positive correlation between CodeBLEU and two metrics, patch compilability (Person's $r=0.22$) and patch plausibility (Person's $r=0.15$), suggesting that CodeBLEU alone is a weak predictor 
for RTT-generated patches.

\begin{figure}
  \centering
  \includegraphics[width=0.45\columnwidth]{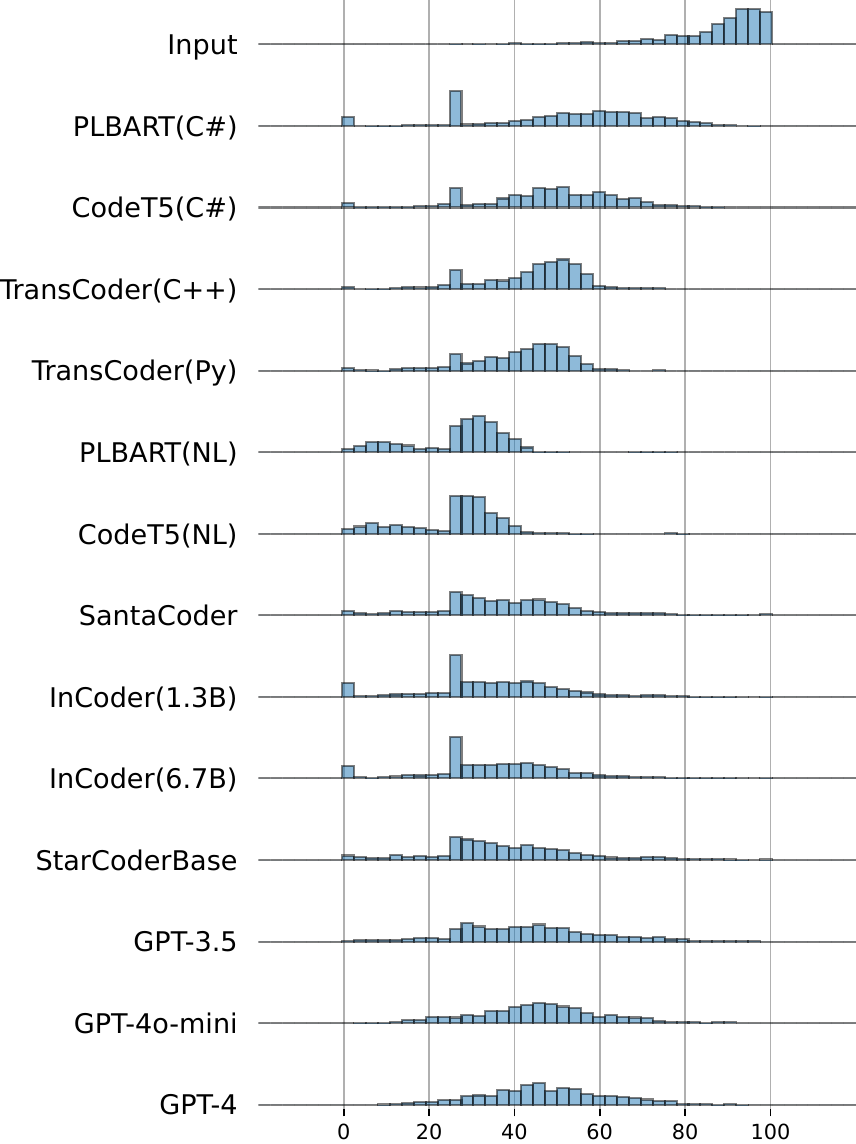}
  \caption{Histogram on the CodeBLEU scores of candidate patches generated with RTT through PL or NL (NL is default, if not mentioned). The distribution is calculated over all patches generated for four APR benchmarks.}
  \label{fig:histplot-codebleu}
  \Description[Comparison of CodeBLEU score distributions across models generating candidate patches.]{The figure presents horizontal histograms showing the distribution of CodeBLEU scores for candidate patches generated by a range of models. 
  The top row shows the distribution for the original input patches, which are skewed toward high scores. 
  Most models produce distributions with peaks around mid-range scores, generally between 30 and 60, indicating moderate similarity to reference patches.}
\end{figure}

\subsubsection{Other common APR metrics}
Exact Match and
BLEU
are frequently used to check whether candidate patches resemble the ground truth in benchmarks~\cite{lu2021:codexglue}.
Since RTT is aimed at finding functionally correct patches, not stylistically equivalent ones, we have found these metrics non-descriptive for RTT.
Especially when using NL as intermediate, RTT can freely deviate from the original buggy code's style, 
evidenced by the average BLEU and CodeBLEU scores ± std over the patches that pass all tests being less than 40.1 ± 0.09 and 63.7 ± 6.7, respectively.

\subsubsection{Limitations of RTT for APR}
Studies show that the original style of writing texts can be diluted by generative language models~\cite{pedreschi2023:social, grimaldi2023:ai}. 
Furthermore, LLMs have been known to generate code containing security flaws~\cite{pearce2022:asleep, tihanyi2023:formai}.
These flaws may compromise the integrity of the application. 
Therefore, it is recommended that developers thoroughly audit and review any generated code.
Our experiments show that bugs in code can be corrected via RTT, 
but we also see that the original styling of the code is not always retained.
Such a restyle leads to challenges with code maintainability, 
which can reduce the willingness of developers to adopt the approach.
This issue will be less of a challenge in projects that enforce a uniform coding style through automated tools.
Moreover, the impact will be lower if RTT is applied in smaller contexts, 
for example, in highly modular projects with localized bugs, where restyling will have limited impact on maintainability. 

\begin{mdframed}[style=mystyle]
\head{Answer to RQ4 (Quantitative Analysis of RTT patches):}
We observe that
\emph{(a)} RTT-generated patches, even if compilable, rarely pass most of the tests, therefore needing significant refinement;
\emph{(b)} RTT can change the code considerably, reducing usefulness of metrics for ground-truth matching, such as BLEU and CodeBLEU;
\emph{(c)} Because RTT can dilute the coding style, it is best used in circumstances where rephrasing does not impact maintainability. 
\end{mdframed} 

\subsection{Qualitative Analysis of Generated Candidate Patches}
We manually analyze patches generated by RTT to find out more about the quality of solutions, how they differ among patches, and what the common flaws are. 
To make the analysis concise, we have chosen to go through HumanEval-Java patches, starting from the problems solved in this work and not solved by other APR approaches in the scope of our comparison (see Section~\ref{sec:comparison-with-previous-work}). This section also includes error and success analysis of the problems in QuixBugs and HumanEval-Java, looking into the reasons for patches to not compile, and the ones that compiled or passed all the tests more frequently than for other problems in the dataset.  

\head{Unique problems solved by RTT:} In HumanEval-Java, some problems were solved uniquely by RTT and not solved by models reported by Jiang et al.~\cite{jiang2023:impact}. 
Several problems have RTT patches with varying efficiency and general logic of algorithms. 
For example, two factorization problems were solved only by models using RTT and not their fine-tuned or base counterparts.
The first problem consists of factorizing a number into the list of its prime factors (FACTORIZE, problem 25), and the second one aims to find the largest of them (LARGEST\_PRIME\_FACTOR, problem 59).
Solutions for FACTORIZE resulted from RTT differ in the way they specify the limit for looping over the numbers: until and including the input number $n$, e.g., SantaCoder's solution in Listing~\ref{lst:factorize-target}, or until its square root, e.g., target code from the dataset and Incoder-6B solution.
An implementation that stops the loop at the square root of the input number is more optimal because it involves fewer iterations. 
In case the iterations stop at the square root, the additional check ($if \; (n > 1)$) for including the number itself is needed, for example, for a prime number 11 as input. 
For reference, the target solution checks for $(n > 1)$ while Incoder-6B --- for $n\ !{=}\ 1$ after the loop. 

\begin{figure}[t]
\setcounter{figure}{0}
\captionsetup{name=Listing}
\centering
\begin{subfigure}[b]{.47\textwidth}
\begin{lstlisting}[language=Java, basicstyle=\scriptsize, frame=tlrb]
public static List<Integer> factorize(int n){
    List<Integer> result = new ArrayList<Integer>();

    int i = 2;
    while (i <= (int)(Math.sqrt(n) + 1)){
        if (n %
            result.add(i);
            n = n / i;
        } else{
            i += 1;
        }
    }
    // bug: the following if-statement was missing
    if (n > 1){
        result.add(n);
    }
    return result;
}
\end{lstlisting}
\caption{Target code, reproduced by Incoder-6B up to renaming of the variables and a small change of if-condition.}
\label{lst:factorize-target}
\end{subfigure}
\hfill %
\begin{subfigure}[b]{.47\textwidth}
\begin{lstlisting}[language=Java, basicstyle=\scriptsize, frame=tlrb]
public static List < Integer > factorize(int n) {
  List < Integer > list = new ArrayList < > ();
  if (n < 0) {
    return list;
  }
  for (int i = 2; i <= n; i++) {
    while (n %
      list.add(i);
      n /= i;
    }
  }
  return list;
}
\end{lstlisting}
\caption{SantaCoder (NL) patch \newline}
\label{lst:factorize-patch}
\end{subfigure}
\caption{Problem: FACTORIZE (HumanEval-Java).} 
\label{lst:factorize}
\end{figure}

The LARGEST\_PRIME\_FACTOR problem is formatted as a class with two methods: an auxiliary method that checks whether a number is prime and the main method that loops over factors of an input number and chooses the largest prime factor (see Listing~\ref{lst:largest-prime-factor-target}). 
This setup is challenging for RTT because the auxiliary method may get lost in the two-step translation. 
However, some of the generated solutions implicitly ensured that the resulting output was prime. 
For example, the RTT patch by StarCoderBase
loops over numbers from $i=2$ to the input number $n$ and modifies it 
with $n=n/i$ (see Listing~\ref{lst:largest-prime-factor-patch-no-isprime}). 
This process ensures that all the non-prime factors are excluded and the resulting largest factor is prime. 
A similar solution was generated by RTT with other models, too, ending the loop at $n$ or $Math.sqrt(n)$ again. 
Interestingly, StarCoderBase does use the check for $is\_prime()$ in some of its solutions as shown in Listing~\ref{lst:largest-prime-factor-patch-with-isprime}. 
While the input to RTT contains this auxiliary method in the context (see Listing~\ref{lst:largest-prime-factor-rtt-steps}), the intermediate translation does not mention it. 
So, the generated output either ``hallucinated'' (as LLMs sometimes do when they import non-existing libraries or methods~\cite{liu2024:exploring, eghbali2024:dehallucinator}) or copied the method from a similar code snippet the model has ``seen'' before.
This unexpected usage of $is\_prime()$ highlights the data leakage issue from either the benchmark in Python or some textbooks the models might have been trained on. 

\begin{figure}[t]
\centering
\captionsetup{name=Listing} 
\begin{subfigure}[b]{.47\textwidth}
\begin{lstlisting}[language=Java, basicstyle=\scriptsize, frame=tlrb]
public static boolean is_prime(int k) {
    if (k < 2)
        return false;
    for (int i = 2; i < k; i += 1){
        if (k %
            return false;
    }
    return true;
}

public static int largest_prime_factor(int n) {
    int largest = 1;
    for (int j = 2; j < n; j += 1) {
        // buggy line: 
        // if (is_prime(j)){
        if (n %
            largest = j;
        }
    }
    return largest;
}
\end{lstlisting}
\caption{\centering Target code. Method $is\_prime$ serves as context, \newline while method $largest\_prime\_factor$ is the main function.}
\label{lst:largest-prime-factor-target}
\end{subfigure}
\hfill
\begin{subfigure}[b]{.47\textwidth}
\begin{lstlisting}[language=Java, basicstyle=\scriptsize, frame=tlrb]
(*@\textbf{INPUT}@*)
<fim_prefix>

/**
@description <fim_suffix>
*/
public static int largest_prime_factor(int n) {
    int largest = 1;
    for (int j = 2; j < n; j += 1) {
        if (is_prime(j)){
            largest = n %
                }
    }
    return largest;
}
<fim_middle>

(*@\textbf{OUTPUT: INTERMEDIATE TRANSLATION}@*)
/**
@description This method returns 
the largest prime factor of the number passed in.
*/

\end{lstlisting}
\caption{\centering Input for RTT with StarCoderBase \newline and its intermediate translation.}
\label{lst:largest-prime-factor-rtt-steps}
\end{subfigure}
\vfill %
\vspace{10pt}
\begin{subfigure}[b]{.47\textwidth}
\begin{lstlisting}[language=Java, basicstyle=\scriptsize, frame=tlrb]
public static int largest_prime_factor(int n) {
    int largest_prime = 2;
    int i = 2;
    while (i <= n) {
        if (n %
            n = n / i;
            if (i > largest_prime) {
                largest_prime = i;
            }
        } else {
            i++;
        }
    }
    return largest_prime;
}
\end{lstlisting}
\caption{\centering StarCoderBase patch (NL). \newline $is\_prime$ method from the input context is not used.}
\label{lst:largest-prime-factor-patch-no-isprime}
\end{subfigure}
\hfill %
\begin{subfigure}[b]{.47\textwidth}
\begin{lstlisting}[language=Java, basicstyle=\scriptsize, frame=tlrb]
public static int largest_prime_factor(int n) {
  int largest = 1;
  for (int i = 2; i <= n; i++) {
    if (n %
      if (is_prime(i)) {
        largest = i;
      }
    }
  }
  return largest;
}
\end{lstlisting}
\caption{StarCoderBase patch (NL) with $is\_prime$ method.\newline}
\label{lst:largest-prime-factor-patch-with-isprime}
\end{subfigure}
\caption{Problem: LARGEST\_PRIME\_FACTOR (HumanEval-Java).} %
\label{lst:largest-prime-factor}
\end{figure}

\head{Differences in Implementation Styles:}
In line with the efficiency discussion, Listing \ref{lst:gcd} refers to the calculation of the greatest common divisor between two numbers.
The proposed solution includes classic recursive calls, while the plausible generated patch shows the iterative version of this recursion.
In this case, the target code is slightly more inefficient due to the recursive calls.
On the other hand, 
Listing \ref{lst:bitcount} shows the classic BIT\_COUNT problem.
The goal is to count how many bits are activated in a given natural number.
There exist not only different approaches for this problem, but their efficiency depends on the context in which they will be used.
The proposed solution follows the Brian Kernighan's algorithm, while the plausible patch generated by several models fosters the readability of the approach.
Although it is true that the target code may be considered slightly more efficient, the difference is negligible in most scenarios.

\begin{figure}[t]
\centering
\captionsetup{name=Listing} 
\begin{subfigure}[b]{.47\textwidth}
\begin{lstlisting}[language=Java, basicstyle=\scriptsize, frame=tlrb, label=lst:solution_gcd]
public static int gcd(int a, int b) {
    if (b == 0) {
        return a;
    }
    else {
        return gcd(b, a%
    }
}
\end{lstlisting}
\caption{Target Code}
\end{subfigure}
\hfill %
\begin{subfigure}[b]{.47\textwidth}
\begin{lstlisting}[language=Java, basicstyle=\scriptsize, frame=tlrb, label=lst:plausible_gcd]
public static int gcd(int a, int b) {
    while (b != 0) {
        int temp = b;
        b = a %
        a = temp;
    }
    return a;
}
\end{lstlisting}
\caption{StarCoderBase patch}
\end{subfigure}
\caption{Problem: GCD (QuixBugs).} %
\label{lst:gcd}
\end{figure}

\begin{figure}[t]
\centering
\captionsetup{name=Listing} 
\begin{subfigure}[b]{.47\textwidth}
\begin{lstlisting}[language=Java, basicstyle=\scriptsize, frame=tlrb, label=lst:solution_bitcount]
public static int bitcount(int n) {
    int count = 0;
    while (n != 0) {
        n = (n & (n - 1));
        count++;
    }
    return count;
}
\end{lstlisting}
\caption{Target Code}
\end{subfigure}
\hfill %
\begin{subfigure}[b]{.47\textwidth}
\begin{lstlisting}[language=Java, basicstyle=\scriptsize, frame=tlrb, label=lst:plausible_bitcount]
public static int bitcount(int n) {
    int count = 0;
    while (n != 0) {
        count += n & 1;
        n >>>= 1;
    }
    return count;
}
\end{lstlisting}
\caption{StarCoderBase patch}
\end{subfigure}
\caption{Problem: BitCount (QuixBugs)} %
\label{lst:bitcount}
\end{figure}

Similar to checking if a number is prime when searching for the largest prime factor, in the problem EVEN\_ODD\_\\PALINDROME (problem 107), the method $is\_palindrome()$ present in the context is sometimes used in the RTT-generated patch and other patches perform this check as a one-liner in the generated code in the if-statement, such as: $if \; (s.equals(new \; StringBuilder(s).reverse().toString())).$ 
Both types of checks happen in several models arbitrarily, so we could not draw any conclusion on correlations with a model size or type.
Noteworthy, in a similar fashion that methods from the input context are not always used in the generated patches (and the presence of those is usually not reflected in the intermediate translation), built-in Java classes are not always used for standard manipulations.
For example, a sorting algorithm can be either reused from a standard Java library or implemented in the patch.  

\head{Common Logical Problems:}
Several problems in HumanEval-Java have method names that have a misleading effect on the outputs of RTT. In this way, CORRECT\_PARENTHESIS (problem 61) has led to one output checking for several types of parenthesis, although the original problem description talks only about the round parentheses ``().'' 
However, we use only the code as input for the first RTT leg that summarizes the code. 
Therefore, we sometimes miss assumptions about the input only present in the original description of the program, for example, those stating that the input will always be composed of round parentheses.
An excessive check for several types of parentheses happens in patches generated for other problems, too. 
This is reflected in the intermediate translation that covered multiple bracket types, even though the input code only had checks for round parentheses. 

The misleading method names are some of the common reasons for low plausibility rates. 
In these cases, an intermediate translation takes the problem far from the input code, and the resulting patch can even have a different number of input arguments, which makes testing on the same dataset's tests impossible.
A similar misleading effect occurred for the solution of  SMALLEST\_CHANGE (problem 73, Listing~\ref{lst:smallest-change}), whereby the model created an intermediate translation that did not describe the intended problem well. 
So, the solution is generated for a different problem based on variable names (coins) and extra variables set up for the solution, but surprisingly, it is correct for the initial problem, too. 
In essence, the only useful check in Listing~\ref{lst:smallest-change-patch} is that $arr[i] \; != \; arr[j]$ which, in this implementation, is realized as $arr[i]  <  arr[j]$. 
Once the condition is true, the counter should be incremented by 1, which is done by $coins \; += \; 1.$
Thus, the variable $min$ is not useful for obtaining the final answer at all.
Moreover, it only makes sense to run the loop as long as $i < j$, which should be an additional condition in the while loop to make the implementation more optimal. 

\begin{figure}[t]
\centering
\captionsetup{name=Listing} 
\centering\begin{subfigure}[b]{.47\textwidth}
\begin{lstlisting}[language=Java, basicstyle=\scriptsize, frame=tlrb]
public static int smallest_change(int[] arr) {
    int result = 0;
    // buggy line: 
    // for (int i = 0; i < arr.length; i += 1) {
    for (int i = 0; i < arr.length / 2; i += 1) {
        if (arr[i] != arr[arr.length - i - 1])
            result += 1;
    }

    return result;
    
}
\end{lstlisting}
\caption{Target code}
\label{lst:smallest-change-target}
\end{subfigure}
\hfill
\begin{subfigure}[b]{.47\textwidth}
\begin{lstlisting}[language=Java, basicstyle=\scriptsize, frame=tlrb]
/**
@description <|mask:0|>
*/
public static int smallest_change(int[] arr) {
    int result = 0;
    for (int i = 0; i < arr.length; i += 1) {
        if (arr[i] != arr[arr.length - i])
                result += 1;
    }
    return result;
}
<|mask:0|>
\end{lstlisting}
\caption{Input for RTT with Incoder-6B}
\end{subfigure}
\hfill %
\begin{subfigure}[b]{.47\textwidth}
\begin{lstlisting}[language=Java, basicstyle=\scriptsize, frame=tlrb]
/**
@description Find the smallest number of coins 
    needed to make change 
    for a given amount of money.

@param: arr - an array of integers representing coins

@return: the smallest number of coins 
    needed to make change 
    for the given amount of money

@exception: ArrayIndexOutOfBoundsException 
    - if array is empty
*/

\end{lstlisting}
\caption{Incoder-6B intermediate translation (NL)}
\end{subfigure}
\hfill %
\begin{subfigure}[b]{.47\textwidth}
\begin{lstlisting}[language=Java, basicstyle=\scriptsize, frame=tlrb]
public static int smallest_change(int[] arr) {
     int min = Integer.MAX_VALUE;
     int coins = 0;
     int n = arr.length;
     int i = 0;
     int j = n-1;
     while(i < n && j >= 0) {
         if(arr[i] < arr[j]) {
             min = Math.min(min, arr[i] + arr[j]);
             coins += 1;
         }
         i++;
         j--;
     }
     return coins;
}
\end{lstlisting}
\caption{Incoder-6B patch}
\label{lst:smallest-change-patch}
\end{subfigure}
\caption{Problem: SMALLEST\_CHANGE (HumanEval-Java).} %
\label{lst:smallest-change}
\end{figure}

\begin{figure}[t]
\centering
\captionsetup{name=Listing} 
\begin{subfigure}[b]{.47\textwidth}
\begin{lstlisting}[language=Java, basicstyle=\scriptsize, frame=tlrb]
public static String concatenate(String[] strings){
    String result = "";
    for (String string : strings) 
        result += string;
    return result;
}
\end{lstlisting}
\caption{Target code}
\end{subfigure}
\hfill %
\begin{subfigure}[b]{.47\textwidth}
\begin{lstlisting}[language=Java, basicstyle=\scriptsize, frame=tlrb]
public static String concatenate (String... strings){
    StringBuilder builder = new StringBuilder();
    for (String s : strings){
        builder.append(s);
    }
    return builder.toString();
}
\end{lstlisting}
\caption{Plausible patch by PLBART (NL)}
\end{subfigure}
\caption{Problem: CONCATENATE (HumanEval-Java).} %
\label{lst:concatenate}
\end{figure}

\head{Frequent Failure and Success Cases:}
Problems with string manipulation usually have high compilability and plausibility rates, even though solution approaches vary. 
The variation occurs both in the Java class used by a patch, e.g., StringBuilder, StringBuffer, or just String, and in the logic for solving the problem. 
Listing \ref{lst:concatenate} displays the concatenation problem in the HumanEval-Java benchmark.
The purpose of the given function is to concatenate strings given an array.
The plausible patch generated by PLBART is similar to the proposed solution.

Continuing the discussion on tasks in HumanEval-Java with string manipulation, in MATCH\_PARENS (problem 119), the bug consisted in the wrong if-statement or, otherwise, wrong place to increment and decrement the counter $val$ of unmatched opening parentheses.
So, the patch should have either swapped the increment and decrement or changed the if-statement $if \; (s.charAt(i) \; == \; '(')$ to  $if \; (s.charAt(i) \; == \; ')').$
One of those changes was present in the correctly generated patches. 
However, a solution created by RTT with SantaCoder used an integer array that served as a stack (see Listing~\ref{lst:match-parens}).
In this solution, opening parentheses are put into a stack that is implemented as an array of integers. 
Index $i$ loops over the input string. 
The variable $open$ updates the index of the latest open and unmatched parenthesis occurring in the input, i.e., at the top of the stack.
When a closing parenthesis occurs in the input string, an element is ``popped'' from the stack by decrementing $open.$ 
The solution can be simplified by manipulating just the $open$ variable (as it happens to $val$ in the target code) and thus have $O(1)$ space instead of $O(n)$ with an extra $stack$ array of size $n$. 
A surprising fact is that the solution with a stack does not use the standard $java.util.Stack$ class.

\begin{figure}[t]
\centering
\captionsetup{name=Listing} 
\begin{subfigure}[b]{.47\textwidth}
\begin{lstlisting}[language=Java, basicstyle=\scriptsize, frame=tlrb]
public static boolean check(String s) {
    int val = 0;
    for(int i = 0; i < s.length(); i++) {
    
        // buggy lines with swapped val++ and val--
        // if (s.charAt(i) == '(') val--;
        //    else val++;
        //    if(val < 0) return false;
        
        if(s.charAt(i) == '(') val++;
        else val--;
        if(val < 0) return false;
    }
    return val == 0;
}
\end{lstlisting}
\caption{Target code}
\end{subfigure}
\hfill %
\begin{subfigure}[b]{.47\textwidth}
\begin{lstlisting}[language=Java, basicstyle=\scriptsize, frame=tlrb]
public static boolean check(String s) {
  int[] stack = new int[s.length()];
  int open = 0;
  for (int i = 0; i < s.length(); i++) {
    if (s.charAt(i) == '(') {
      stack[++open] = i;
    } else if (s.charAt(i) == ')') {
      if (open == 0) {
        return false;
      }
      stack[--open] = i;
    }
  }
  return open == 0;
}
\end{lstlisting}
\caption{\centering SantaCoder patch (NL)}
\end{subfigure}
\caption{Problem: MATCH\_PARENS (HumanEval-Java). } %
\label{lst:match-parens}
\end{figure}

One source for common errors is the challenge of dealing with linked lists and nodes. 
Thus, patches for problems that require node processing usually fail to compile because of the wrong node handling, such as calling the $next()$ method instead of $getSuccessor().$
Some general observations are that smaller models, specifically both Incoders, sometimes output repetitive tokens or very similar lines of code, such as an if-condition that compares a variable to all numbers from 1 to 50 one by one. 
This can happen when a model's capacity to train is not fully used, and performance can benefit from more training.
Incoder models also more frequently generate several alternatives within one generation, which looks more like a batch generation, but instead of having several outputs in a batch (or a list), all of them are appended to one generated string.  

Some of the most frequently solved problems (with True Positive Rate=1 for the most generated outputs and the majority of models) are common programming or mathematical tasks for which there should be plenty of training data available. 
Such problems are, for example, GCD and BITCOUNT in QuixBugs, and Fibonacci sequence, string concatenation, and again GCD calculation in HumanEval-Java.
Many output patches are similar to the target solutions presented in the datasets up to renaming the variables, changing the order of conditional statements, or other operations that do not affect the logic of the solutions. 
These observations confirm that even though some of the datasets possibly leak to the training set of the models, the models do not reproduce this code line by line.

\begin{mdframed}[style=mystyle]
\head{Answer to RQ5 (Qualitative analysis of RTT patches):}
We find that
\emph{(a)} While many HumanEval-Java patches are very similar to target code, they have different variable names, and there exist other solutions that do not resemble target code at all, which confirms that even if the models were exposed to the benchmarks, they did not memorize them to the point of copy-pasting the answers;
\emph{(b)} Using the code from the surrounding problem context is challenging for RTT patches, and some simple checks made in the surrounding contextual methods are sometimes replaced by one-line checks in the patch body;
\emph{(c)} Time and space complexity vary across the generated patches, given the fact that the prompts did not urge models to optimize for those;
\emph{(d)} In RTT, method names can sometimes be misleading and result in a shift to a different problem in the intermediate translation;
\emph{(e)} RTT as an APR technique results in an indirect fix to a problem leading to more creative solutions as opposed to direct line-specific fixes of fine-tuned models.
\end{mdframed} 

\subsection{Impact and Potential Usage of RTT in Code Repair Frameworks}

The design and development of a single tool or methodology to address the entire spectrum of software bugs can be challenging~\cite{lutellier2020:coconut}.
For this reason, the evolution of APR has emphasized the value of ensemble approaches~\cite{kang2022:language}.
In ensemble frameworks, diverse tools and methodologies contribute their unique strengths in identifying and fixing software bugs.
These ensemble approaches underscore the importance of the diversity of tools and align with realistic software maintenance scenarios, where a combination of techniques is employed to achieve the best results.

In the context of APR, agent-based approaches can be used to exemplify this ensemble approach.
They
are expected to become a new standard for decision-making support and coding in the future. 
In such systems, one agent can play the role of a router and other agents suggest alternative solutions for a task under consideration, together composing an ensemble setting for generating a solution~\cite{shen2023:hugginggpt}. 
In this case, it is beneficial to the final repair target if each individual agent proposes good-quality solutions that differ from the solutions suggested by other agents. 

RTT can be viewed as a complementary approach for a repair agent in an APR agent-based framework, where some agents can apply self-reflection or self-debugging, others apply NMT approaches to predict repairing code edits, a cloze-style repair agent fills in the masked buggy lines with a suggestion for correct lines, and an RTT agent generates an alternative repair proposal by translating buggy code to another language and back. 
Because RTT repairs 46 unique problems not solved by models that employ cloze-style repair, its added value in a multi-agent code repair environment can boost the performance of the framework. 
Future work can explore the avenue of creating multi-agent APR frameworks based on language models trained for code with tools such as LangChain,\footnote{https://www.langchain.com/} LlamaIndex\footnote{https://www.llamaindex.ai/} or AutoGPT.\footnote{https://github.com/Significant-Gravitas/AutoGPT}

\section{Threats to Validity}%
This section discusses four types of threats to validity for this study, structured cf. Wohlin et al.~\cite[Sec. 6.7 \& 6.8]{wohlin2000:experimentation}. 

\head{Internal Validity:}
To support the validity of our results, we applied RTT 
with two intermediate representations
to four benchmarks and tested nine models. 
As the benchmarks are publicly available, 
there is a risk that they were used during training, 
also referred to as \emph{data leakage}. 
This threat can be mitigated by 
using models that remove the benchmarks from their training data.
Here, we use HumanEval-Java, which was constructed after the training of any of the models used in this work (except GPT-4o-mini).
For the other three datasets, 
we find an exact match in only 0.03\% of the generated candidate patches, 
which reinforces the validity of the data and results by ensuring that even if the models were trained on the used benchmarks, they failed to copy-paste patches.

\head{Construct Validity:}
To evaluate the RTT performance,
we apply widely used APR metrics, including compilability and plausibility rates.
For metrics that depend on test suites, low-quality or easy-to-pass tests could positively bias the evaluation.
We mitigate this risk by employing four widely used APR benchmarks with different bug types.
However, most of our evaluation uses plausibility as a proxy for correctness, which comes with inherent limitations, such as ignoring potential overfitting of patches to the test code.
To address this threat, we manually assess the correctness of over 5,000 patches generated for HumanEval-Java.
This helps
to balance the difficulty of problems covered by the benchmark 
with
the resources required to evaluate multiple types of patches for each problem and each LLM. 
Furthermore, HumanEval-Java was created to minimize the effect of possible data leakage, while QuixBugs and Defects4J were already available online.
Finally, evaluating the correctness of a large number of patches for Defects4J would require an unfeasible amount of resources 
and would be prone to evaluator (human) errors, since many of the bugs require in-depth knowledge of the projects in Defects4J.

\head{External Validity:}
Threats to the external validity 
concern the 
generalizability of our approach.
We have validated the approach on a representative sample of APR benchmarks, but have not extended the results to language pairs that were not covered. 
However, we focused our evaluation on single-hunk bugs, therefore, effectiveness may not transfer to more complex multi-hunk or multi-file bugs.
These multi-hunk bugs may represent a more realistic scenario that requires richer context and evaluation of cross-hunk consistency.
Moreover, we applied the approach using only nine transformer-based models. 
Extending the evaluation requires more computational resources and language models that comply with the RTT requirements in Section~\ref{sec:models}.

\head{Conclusion Validity:}
For our experiments, we used off-the-shelf language models that are publicly available and can be used without retraining. 
The four benchmarks are also publicly available and widely used in APR research.
To support open science and enable replication and verification of our work, 
a replication package is available.\footref{replication}%

\section{Directions for Further Research}
\label{sec:future}

\noindent
This study opens up several avenues for future research, such as expansion of key components of the current study by adding new models, benchmarks, or metrics.
Our replication package can be used as a base to expand the study to new models, intermediate languages, and datasets.
Interesting directions for future work include:

\head{Model Variation:} Investigating models with the same architecture but varied size, e.g. CodeT5+~\cite{wang2023:codet5}, may shed light on whether some capabilities show only after a specific size threshold~\cite{kaplan2020:scaling, wei2022:emergent}. 
This exploration can result in some conclusions on the scalability of RTT and how model complexity may impact precision and quality of repairs.
The emergence of unique repair capabilities at certain model size thresholds can reveal insights on the optimization of APR tools' performance.
Another straightforward variation that can influence repair performance is usage of different models in each RTT leg. For example, choosing the newest best performing model for summarization and the best performing one for code generation has a potential on improving RTT via natural language.
\head{Cross-entropy Analysis:} A more granular analysis of cross-entropy along with increased sampling from models can provide insights into the naturalness hypothesis and its effects on the approach~\cite{xia2023:automated}.
Studying how similar the probability distribution of repaired code match natural, bug-free code can help researchers assess the effectiveness of APR methods.
This could result in the development of new loss functions that could capture nuances of successful code repair.
\head{Extend Intermediate Representations:} Researching the properties of other intermediate representations, not only other programming languages, but also pseudocode or combinations of PL and NL.
This exploration can identify what intermediate representations are the most effective for each programming language or bug type.
As a result, the use of specifically chosen intermediate can provide with customized solutions that take advantage of the strengths of the various coding languages.
\head{Error Analysis:} Classifying the bugs from the benchmark into groups that share similar bug properties could provide trends about which type of errors are fixed with some specific intermediate representations.
This information could also be used to generate prompts for the problems adapted to failing tests or compilation errors, similarly to \citet{liventsev2023:fully}.
Furthermore, knowing which leg of the round-trip can provide insights for experiments that combine multiple models. Gaining a deeper understanding of the characteristics and limitations of unique patches, such as model rigidity or optimal number of outputs, can help develop more sophisticated translation and repair techniques.
\head{Multiple Round Trips:} Since the process relies on the regression toward the mean, repeated RTT may increase the bug-fixing behavior.
This investigation can lead to a more nuanced understanding into the mean of the code learned by language models.
Researching if repeated translations can incrementally refine the repairs may lead to more accurate and robust results, enhancing the bug-fixing capabilities of RTT-based APR approaches.
\head{Granularity of RTT:}
One can restrict RTT from translating the whole function to applying changes to certain masked areas of the code. 
However, RTT's bug fixing capabilities rely on context to accurately filter out bugs.
Consequently, future research can investigate if an optimal granularity exists that balances accurate modifications with the amount of context needed.
Moreover, studying the effect of increasing the context could allow RTT to be applied to more realistic multi-hunk and multi-file bugs where a rich context is available but complexity is also increased.

\section{Conclusion}
\label{sec:conclusions}

In this work, we explore the latent capability of LLMs for automated program repair illuminated by the round-trip translation through an intermediate representation.
Our experiments confirm the viability of RTT for APR, but also show potential pitfalls and limitations.
Although RTT does not outperform state-of-the-art NMT and cloze-style fine-tuned models for APR, 
we find that RTT, without any fine-tuning costs, 
is able to repair 46 unique bugs that were not fixed by the same LLMs after fine-tuning them for NMT and cloze-type APR.
In more detail, we find that RTT through NL (English) 
as an intermediate translation generated plausible patches for 100 of 164 bugs with GPT-4 on the HumanEval-Java benchmark, and 97 were determined to be correct in our manual assessment.
When using PL as intermediary, we obtain two main insights:
First, the intermediate representation should be sufficiently distinct from the original language to ensure effective rectification.
Secondly, larger LLMs consistently provide better results than their smaller counterparts.

Through close examination, we uncover several properties and characteristics of RTT-generated patches.
The inherent nature of the approach to introduce substantial changes to the code makes it inefficient to use common ground-truth matching metrics, such as BLEU, to assess efficacy.
Our qualitative analysis of the patches generated for QuixBugs and HumanEval-Java highlights the complex relationship between RTT and the naturalness of code.
While RTT leverages naturalness and the regression toward the mean to remove bugs, 
this process may also dilute the original code author's style, remove comments, and substantially reformat the code. 
Therefore, RTT is better suited in circumstances where such rephrasing does not impact maintainability.
Together with the ability to provide fixes for bugs that other approaches do not solve, 
this makes RTT a technique that is particularly useful in collaborative scenarios where multiple methods are used to address a task, 
for example, in multi-agent or ensemble frameworks.
Not only can RTT expand the repertoire of available tools, it also introduces a novel dimension to patch generation by applying a non-traditional approach.
Moreover, this expansion comes at a low cost because the approach does not require any fine-tuning of LLMS for the APR task.

Finally, we have sketched a number of promising directions for future research, 
including experimentation with a wider range of models, a more granular analysis of cross-entropy in code, 
investigating properties of other intermediate representations, a qualitative analysis of errors, 
experimenting with multiple round-trips, and restricting RTT to certain `masked' areas of the code. 
We believe that new insights into any of these areas will help further the application of RTT to make software engineers be even better at their jobs.

\begin{acks}
This work is supported by the Research Council of Norway through the secureIT project (IKTPLUSS \#288787), and by the European Union the Horizon Europe Marie Sk\l{}odowska-Curie Actions (\#101151798).
The empirical evaluation presented in this article made use of the Experimental Infrastructure for Exploration of Exascale Computing (eX3), 
financially supported by the Research Council of Norway under contract \#270053. 
In addition, we acknowledge Sigma2, Norway for awarding this project access to the LUMI supercomputer, owned by the EuroHPC Joint Undertaking, hosted by CSC (Finland) and the LUMI consortium through the Research Council of Norway. 
This research builds on work that has been submitted in partial fulfillment of the requirements for the Master’s degree of the first author at KTH, Sweden.
\end{acks}

\bibliographystyle{ACM-Reference-Format}
\begin{DIFnomarkup} %
\bibliography{RTT}
\end{DIFnomarkup} %

\end{document}